\documentclass{amsart}
\usepackage{amsfonts}

\usepackage{amscd}

\newtheorem{theorem}{Theorem}
\theoremstyle{plain}

\newtheorem{definition}{Definition}

\newtheorem{proposition}{Proposition}

\numberwithin{equation}{section}

\input{tcilatex}

\begin{document}
\title{ON\ ENTANGLED INFORMATION AND QUANTUM\ CAPACITY.}
\author{Viacheslav P Belavkin}
\address{Department of Mathematics, University of Nottingham, NG7 2RD Nottingham, UK}
\date{July 2, 2000}
\subjclass{Quantum Information}
\keywords{Entanglements, Compound States, Dimensional Entropy and Quantum Information. }
\thanks{This work was partially supported by the Royal Society grant for UK--Japan
collaboration.}
\maketitle

\begin{abstract}
The pure quantum entanglement is generalized to the case of mixed compound
states on an operator algebra to include the classical and quantum encodings
as particular cases. The true quantum entanglements are characterized by
quantum couplings which are described as transpose-CP, but not Completely
Positive (CP), trace-normalized linear positive maps of the algebra.

The entangled (total) information is defined in this paper as a relative
entropy of the conditional (the derivative of the compound state with
respect to the input) and the unconditional output states. Thus defined the
total information of the entangled states leads to two different types of
the entropy for a given quantum state: the von Neumann entropy, or
c-entropy, which is achieved as the supremum of the information over all
c-entanglements and thus is semi-classical, and the true quantum entropy, or
q-entropy, \ which is achieved at the standard entanglement. The q-capacity,
defined as the supremum over all entanglements, coincides with the
topological entropy. In the case of the simple algebra it doubles the
c-capacity, coinciding with the rank-entropy. The conditional q-entropy
based on the q-entropy, is positive, unlike the von Neumann conditional
entropy, and the q-information of a quantum channel is proved to be additive.
\end{abstract}

\section{Introduction}

Quantum theory, which celebrated its 100 years anniversary last December,
gives new possibilities for transmission of information which cannot be
explained in the framework of information theory based on the classical
(Kolmogorovian) probability theory. These possibilities are due to the \emph{%
entanglements}, the specifically quantum (q-) correlations which were first
studied by Schr\"{o}dinger who introduced this term in his analysis of EPR
paradox (for more details of this history see the anniversary review paper 
\cite{Be00d}).

Many authors have recently suggested to use the entanglements for quantum
information processes in quantum computation, quantum teleportation, and
quantum cryptography \cite{Ben93, Sch93, JoSch94}. The mathematical study of
entanglement as a special type of quantum correlations from an operational
point of view has been initiated in \cite{Be_Ohy98, Be_Ohy00}. In these
papers the entangled mutual information was introduced as the von Neumann
entropy of the entangled compound state related to the product of marginal
states in the sense of Lindblad, Araki and Umegaki relative entropy \cite
{Lin73, Ara76, Ume}. The corresponding quantum mutual information leads to
an entropy bound for quantum capacity, the additivity of which is not
obvious for non-trivial quantum channels.

In this paper we will use another possibility to define the entangled mutual
entropy, based on the alternative definition of relative entropy first
introduced in \cite{Be_Sta82, Be_Sta84}. As it was proved in \cite{OP93} our
relative entropy is larger than the LAU relative entropy, so that based on
it quantum mutual information will give a larger entropy bound for quantum
capacity.

We are going to prove that this bound for quantum capacity is in a sense
additive, so that there is no need to consider under certain conditions this
mutual information for the powers of quantum channel in order to guarantee
that this entropy bound gives the real upper bound for long quantum block
encodings. As far as we know this is first such measure of quantum capacity,
although it is larger of \cite{Be_Ohy98, Be_Ohy00} and all other earlier
suggested measures.

For the benefit of reader we repeat all needed definitions and notations
related to the entanglement from \cite{Be_Ohy98, Be_Ohy00} in the first part
of this paper. As in these papers we shall use the word entanglement in the
generalized sense including the classical (c-) correlations as
c-entanglements, and calling non-classical correlations as true quantum
entanglements.

We shall show that any compound state can be achieved by a generalized
entanglement, and the classically (c-) entangled states of c-q encodings and
q-c decodings can be achieved by d-entanglements, the diagonal
c-entanglements for these disentangled states. The pure orthogonal
disentangled compound states are most informative among the c-entangled
states in the sense that the maximum of mutual information over all
c-entanglements is achieved on the extreme c-entangled states as the von
Neumann entropy \QTR{cal}{\textsf{S}}$\left( \varsigma \right) $ of a given
normal state $\varsigma $. Thus the maximum of mutual entropy over all
classical couplings, described by c-entanglements of a classical probe
systems $\mathcal{A}$ to the system $\mathcal{B}$, is bounded by the
c-capacity $\mathsf{C}=\log \mathrm{rank}\mathcal{B}$, where $\mathrm{rank}%
\mathcal{B}$ is the dimensionality of a maximal Abelian subalgebra $\mathcal{%
A}\subset \mathcal{B}$.

We prove that the truly entangled states are most informative in the sense
that the maximum of mutual entropy over all entanglements to the quantum
system $\mathcal{B}$ is achieved by an extreme entanglement of the probe
system $\mathcal{A}=\mathcal{B}$, called standard for a given $\varsigma $.
The mutual information for such extreme q-compound state defines another
type of entropy, the q-entropy $\mathsf{H}\left( \varsigma \right) \leq 2$%
\QTR{cal}{\textsf{S}}$\left( \varsigma \right) $. The maximum of mutual
entropy over all quantum couplings, described by true quantum entanglements
of probe systems $\mathcal{A}$ to the system $\mathcal{B}$ is bounded by $%
\mathsf{C}_{q}=\log \dim \mathcal{B}$.

In this paper we consider the case of a discrete decomposable W*-algebra $%
\mathcal{B}$ for which the results are achieved by relatively simple proofs.
The purely quantum case of a simple algebra $\mathcal{B}=\mathcal{L}\left( 
\mathcal{H}\right) $, for which some proofs are rather obvious, will be also
published elsewhere.

\section{Compound States and Entanglements}

Let $\mathcal{H}$ denote a separable Hilbert space of quantum system, and $%
\mathcal{L}\left( \mathcal{H}\right) $ be the algebra of all linear
operators $B:\mathcal{H}\rightarrow \mathcal{H}$ having the Hermitian
adjoints $B^{\dagger }$ on $\mathcal{H}$. In order to include the classical
discrete systems as a particular quantum case, we shall fix a decomposable
subalgebra $\mathcal{B}\subseteq \mathcal{L}\left( \mathcal{H}\right) $ of
bounded observables $B\in \mathcal{B}$ of the block-diagonal form $B=\left[
B\left( i\right) \delta _{i}^{k}\right] $, where $B\left( i\right) \in 
\mathcal{L}\left( \mathcal{H}_{i}\right) $ are arbitrary bounded operators
in Hilbert subspaces $\mathcal{H}_{i}$ corresponding to an orthogonal
decomposition $\mathcal{H}=\oplus _{i}\mathcal{H}_{i}$.

A normal state on $\mathcal{B}$ is a positive linear functional $\varsigma :%
\mathcal{B}\rightarrow \mathbb{C}$ which can be expressed as 
\begin{equation}
\varsigma \left( B\right) =\mathrm{Tr}_{\mathcal{G}}\chi ^{\dagger }B\chi =%
\mathrm{Tr}B\sigma ,\text{ \quad }B\in \mathcal{B}\text{.}  \label{1.1}
\end{equation}
Here $\mathcal{G}$ is another separable Hilbert space, $\chi $ is a
Hilbert-Schmidt operator from $\mathcal{G}$ to $\mathcal{H}$, $\chi
^{\dagger }$ is its adjoint $\mathcal{H}\rightarrow \mathcal{G}$, and $%
\mathrm{Tr}_{\mathcal{G}}$ (or simply $\mathrm{Tr}$ if there is no
ambiguity) denotes trace in $\mathcal{G}$ (or in $\mathcal{H}$). This $\chi $
is called the amplitude operator (just amplitude if $\mathcal{G}$ is one
dimensional space $\mathbb{C}$, $\chi =\psi \in \mathcal{H}$ with $\chi
^{\dagger }\chi =\Vert \psi \Vert ^{2}=1$ in which case $\chi ^{\dagger }$
is the functional $\psi ^{\dagger }$ from $\mathcal{H}$ to $\mathbb{C)}$. If
the normal state $\varsigma $ is pure on the decomposable algebra $\mathcal{B%
}$, then the density operator $\sigma =\chi \chi ^{\dagger }$ is uniquely
defined as one dimensional projector $P=\psi \psi ^{\dagger }\in \mathcal{B}$%
. For mixed states the $\sigma $ in (\ref{1.1}) may not be unique, but it is
uniquely defined as a positive trace one operator ($\sigma \geq 0$, $\mathrm{%
Tr}\sigma =1$) by an additional condition $\sigma \in \mathcal{B}$, and is
called in this case the \emph{probability operator on} $\mathcal{B}$,
denoted as $\sigma =$\textrm{P}$_{\mathcal{B}}$.

The amplitude operator is not unique, however it is defined uniquely up to a
unitary transform $\chi ^{\dagger }\mapsto U\chi ^{\dagger }$ in $\mathcal{G}
$ as a \emph{probability amplitude} by the additional condition $\chi \chi
^{\dagger }\in \mathcal{B}$. Such a $\chi $ always exists as square root of
the decomposable probability operator $\mathrm{P}_{\mathcal{B}}=\oplus 
\mathrm{P}_{\mathcal{B}}\left( i\right) \in \mathcal{B}$ with the components 
\textrm{P}$_{\mathcal{B}}\left( i\right) \in \mathcal{L}\left( \mathcal{H}%
_{i}\right) $ normalized as 
\begin{equation*}
p\left( i\right) =\mathrm{Tr}_{\mathcal{H}_{i}}\mathrm{P}_{\mathcal{B}%
}\left( i\right) \geq 0,\quad \sum_{i}p\left( i\right) =1.
\end{equation*}

We denote by $\mathcal{B}_{\ast }\subset \mathcal{B}$ the predual space to $%
\mathcal{B}$ identified with the Banach subspace $\mathcal{T}\left( \mathcal{%
H}\right) \cap \mathcal{B}=\oplus \mathcal{T}\left( \mathcal{H}_{i}\right) $
of trace class operators $\sigma =\oplus \sigma _{i}$, where $\sigma _{i}\in 
\mathcal{T}\left( \mathcal{H}_{i}\right) $. The probability operators (the
unique densities) $\mathrm{P}_{\mathcal{A}}$, $\mathrm{P}_{\mathcal{B}}$ of
the states $\varrho $, $\varsigma $ on different algebras $\mathcal{A}%
\subseteq \mathcal{L}\left( \mathcal{G}\right) $, $\mathcal{B}\subseteq 
\mathcal{L}\left( \mathcal{H}\right) $ will be usually denoted by the
variables $\rho \in \mathcal{A}_{\ast }$ and $\sigma \in \mathcal{B}_{\ast }$
corresponding to their Greek variations $\varrho $ and $\varsigma $.

In general, $\mathcal{G}$ is not one dimensional, the dimensionality $\dim 
\mathcal{G}$ must not be less than $\mathrm{rank}\rho $, the dimensionality
of the range $\mathrm{ran}\chi ^{\dagger }$ of the density operator $\tilde{%
\rho}=\chi ^{\dagger }\chi $ coinciding with $\mathrm{rank}\sigma $ of the
probability operator $\sigma =\chi \chi ^{\dagger }$. This implies that the $%
\mathrm{rank}\mathcal{A}$ of any discretely decomposable subalgebra $%
\mathcal{A}\subseteq \mathcal{L}\left( \mathcal{G}\right) $ having $\rho $
as the probability operator $\rho \in \mathcal{A}$ must not be less than $%
\mathrm{rank}\sigma $ if $\sigma =\chi \chi ^{\dagger }$.

We can always equip $\mathcal{H}$ (and we will equip the other Hilbert
spaces) with an isometric involution $J=J^{\dagger }$, $J^{2}=I$, having the
properties of complex conjugation on $\mathcal{H}$, 
\begin{equation*}
J\sum \lambda _{j}\eta _{j}=\sum \bar{\lambda _{j}}J\eta _{j},\quad \forall
\lambda _{j}\in \mathbb{C},\eta _{j}\in \mathcal{H}\text{,}
\end{equation*}
with respect to which the fixed density $\sigma $ is invariant, $J\sigma
J=\sigma $, as a real element of an invariant Abelian subalgebra $\mathcal{A}%
=J\mathcal{A}J\ $of $\mathcal{L}\left( \mathcal{H}\right) $. The latter can
also be expressed as the symmetricity property $\tilde{\varsigma}=\varsigma $
of the state $\varsigma \left( B\right) =\mathrm{Tr}B\sigma $ given by the
real Hermitian and so symmetric density operator $\tilde{\sigma}=\bar{\sigma}
$ on $\mathcal{H}$ with respect to the complex conjugation $\bar{B}=JBJ$ and
the tilde operation (transposition) $\tilde{B}=JB^{\dagger }J$ on $\mathcal{B%
}$. One can always assume that $J$ is the standard complex conjugation in an
eigen-representation of $\sigma $, and take the maximal Abelian subalgebra $%
\mathcal{A}\subset \mathcal{L}\left( \mathcal{H}\right) $ of all diagonal
operators in this basis.

Given the amplitude operator $\chi $, one can define not only the states $%
\varsigma $ by $\sigma =\chi \chi ^{\dagger }$ on the algebra $\mathcal{B}$
but also a \emph{pure compound} state $\varpi $ on the algebra of all
bounded operators on the tensor product Hilbert space $\mathcal{G}\otimes 
\mathcal{H}$ by

\begin{equation*}
\varpi \left( A\otimes B\right) =\mathrm{Tr}_{\mathcal{H}}B\chi \tilde{A}%
\chi ^{\dagger }=\mathrm{Tr}_{\mathcal{G}}\chi ^{\dagger }B\chi \tilde{A}.
\end{equation*}
Thus defined $\varpi $ is uniquely extended by linearity to a normal state
on the algebra $\mathcal{L}\left( \mathcal{G}\otimes \mathcal{H}\right) $
generated by all the linear combinations $C=\sum \lambda _{j}A_{j}\otimes
B_{j}$ due to $\varpi \left( I\otimes I\right) =\mathrm{Tr\,}\chi ^{\dagger
}\chi =1$ and 
\begin{eqnarray*}
\varpi \left( C^{\dagger }C\right) &=&\sum_{i,k}\bar{\lambda}_{i}\lambda _{k}%
\mathrm{Tr}_{\mathcal{G}}\tilde{A}_{k}\tilde{A}_{i}^{\dagger }\chi ^{\dagger
}B_{i}^{\dagger }B_{k}\chi \\
&=&\sum_{i,k}\bar{\lambda}_{i}\lambda _{k}\mathrm{Tr}_{\mathcal{G}}\tilde{A}%
_{i}^{\dagger }\chi ^{\dagger }B_{i}^{\dagger }B_{k}\chi \tilde{A}_{k}=%
\mathrm{Tr}_{\mathcal{G}}X^{\dagger }X\geq 0,
\end{eqnarray*}
where $X=\sum_{j}\lambda _{j}B_{j}\chi \tilde{A}_{j}$.

This compound state $\varpi $ is pure on $\mathcal{L}\left( \mathcal{G}%
\otimes \mathcal{H}\right) $, and it is entangled unless its marginal state $%
\varsigma $ is also pure. Indeed, $\varpi $ corresponds to the amplitude $%
\psi \in \mathcal{G}\otimes \mathcal{H}$ defined by an involution $J$ in $%
\mathcal{G}$ as 
\begin{equation*}
\left( \zeta \otimes \eta \right) ^{\dagger }\psi =\eta ^{\dagger }\chi
J\zeta ,\quad \forall \zeta \in \mathcal{G},\eta \in \mathcal{H}.
\end{equation*}
This definition implies 
\begin{equation*}
\psi ^{\dagger }\left( A\otimes B\right) \psi =\mathrm{Tr}B\chi JA^{\dagger
}J\chi ^{\dagger }\quad \forall A\in \mathcal{L}\left( \mathcal{G}\right)
,B\in \mathcal{L}\left( \mathcal{H}\right)
\end{equation*}
as it can be easily seen for $A=\zeta \zeta ^{\dagger }$, $B=\eta \eta
^{\dagger }$, and $\varsigma $ and $\varrho $ are the marginals of $\varpi $
defined as 
\begin{equation}
\psi ^{\dagger }\left( I\otimes B\right) \psi =\mathrm{Tr}_{\mathcal{H}%
}B\sigma ,\quad \psi ^{\dagger }\left( A\otimes I\right) \psi =\mathrm{Tr}_{%
\mathcal{G}}A\rho .  \label{1.2}
\end{equation}
As follows from the next theorem, any pure compound state 
\begin{equation*}
\varpi \left( A\otimes B\right) =\psi ^{\dagger }\left( A\otimes B\right)
\psi ,\quad A\in \mathcal{A},B\in \mathcal{B}
\end{equation*}
given on the decomposable $\mathcal{A}\otimes \mathcal{B}$ by a probability
amplitude $\psi \in \mathcal{G}\otimes \mathcal{H}$ with $\psi \psi
^{\dagger }\in \mathcal{A}\otimes \mathcal{B}$, can be achieved as described
by a unique entanglement of its marginal states $\varrho $ and $\varsigma $%
.\smallskip

\begin{theorem}
Let $\varpi :\mathcal{A}\otimes \mathcal{B}\rightarrow \mathbb{C}$ be a
compound state 
\begin{equation}
\varpi \left( A\otimes B\right) =\mathrm{Tr}_{\mathcal{F}}\upsilon ^{\dagger
}\left( A\otimes B\right) \upsilon ,  \label{1.3}
\end{equation}
defined by an amplitude operator $\upsilon :\mathcal{F}\rightarrow \mathcal{G%
}\otimes \mathcal{H}$ on a separable Hilbert space $\mathcal{F}$ into the
tensor product Hilbert space $\mathcal{G}\otimes \mathcal{H}$ with 
\begin{equation*}
\upsilon \upsilon ^{\dagger }\in \mathcal{A}\otimes \mathcal{B},\quad 
\mathrm{Tr}_{\mathcal{F}}\upsilon ^{\dagger }\upsilon =1.
\end{equation*}
Then this state is achieved by an entangling operator $\chi :\mathcal{G}%
\rightarrow \mathcal{F}\otimes \mathcal{H}$ as 
\begin{equation}
\varpi \left( A\otimes B\right) =\mathrm{Tr}_{\mathcal{F}\otimes \mathcal{H}%
}\left( I\otimes B\right) \chi \tilde{A}\chi ^{\dagger }=\mathrm{Tr}_{%
\mathcal{G}}\chi ^{\dagger }\left( I\otimes B\right) \chi \tilde{A}
\label{1.4}
\end{equation}
of the states (\ref{1.2}) with $\rho =J\chi ^{\dagger }\chi J$ and $\sigma =%
\mathrm{Tr}_{\mathcal{F}}\chi \chi ^{\dagger }$, where $\chi $ is an
operator $\mathcal{G}\rightarrow \mathcal{F}\otimes \mathcal{H}$ satisfying
the conditions 
\begin{equation*}
\mathrm{Tr}_{\mathcal{F}}\chi \mathcal{A}\chi ^{\dagger }\subset \mathcal{B}%
,\;\chi ^{\dagger }\left( I\otimes \mathcal{B}\right) \chi \subset \mathcal{A%
}.
\end{equation*}
The amplitude operator $\chi $ is uniquely defined by $\tilde{\chi}%
U=\upsilon $, where 
\begin{equation}
\left( \zeta \otimes \eta \right) ^{\dagger }\tilde{\chi}\xi =\left( J\xi
\otimes \eta \right) ^{\dagger }\chi J\zeta ,\quad \forall \xi \in \mathcal{F%
},\zeta \in \mathcal{G},\eta \in \mathcal{H},  \label{1.7}
\end{equation}
up to a unitary transformation $U$ of the minimal space $\mathcal{F}=\mathrm{%
ran}\upsilon ^{\dagger }$ equipped with an isometric involution $J$.
\end{theorem}

\proof%
%
Without loss of generality we can assume that the space $\mathcal{F}$ is a
subspace of $\ell ^{2}\left( \mathbf{N}\right) $ for the diagonal
representation of $\upsilon ^{\dagger }\upsilon $ equipped with the standard
complex conjugation $C$ just as the space $\mathcal{G}$ is in a diagonal
representation of $\chi ^{\dagger }\chi $. In these canonical basises of $%
\mathcal{F}$ and $\mathcal{G}$ the amplitude operator $\chi =\sum \chi
\left( n\right) \langle n|$ can be defined as the block-matrix $\sum
|k\rangle \otimes \chi _{k}\left( n\right) \langle n|$ transposed to $\sum
|n\rangle \otimes \chi _{k}\left( n\right) \langle k|$, where the amplitudes 
$\psi _{k}\left( n\right) \in \mathcal{H}$ are given by the matrix elements $%
\eta ^{\dagger }\chi _{k}\left( n\right) =\left( \langle n|\otimes \eta
^{\dagger }\right) \upsilon |k\rangle $: 
\begin{eqnarray*}
\mathrm{Tr}_{\mathcal{G}}\tilde{A}\chi ^{\dagger }\left( I\otimes B\right)
\chi &=&\sum_{n,m}\left\langle n\right| \tilde{A}\left| m\right\rangle \chi
_{k}^{\dagger }\left( m\right) B\chi _{k}\left( n\right) \\
&=&\sum_{n,m}\chi _{k}^{\dagger }\left( m\right) \left\langle m\right|
A\left| n\right\rangle B\chi _{k}\left( n\right) =\mathrm{Tr}_{\mathcal{F}%
}\upsilon ^{\dagger }\left( A\otimes B\right) \upsilon \text{ .}
\end{eqnarray*}
In any other ortho-normal basis $\left\{ \xi _{k}\right\} \subset \mathcal{F}
$ the involution $J:\mathcal{F}\rightarrow \mathcal{F}$ satisfying $J\xi
_{k}=\xi _{k}$ is defined as $U^{\dagger }CU$, and $\upsilon =\sum |n\rangle
\otimes \psi _{k}\left( n\right) \xi _{k}^{\dagger }=\tilde{\chi}U$, where $%
U=\sum |k\rangle \xi _{k}^{\dagger }$. The isometric transformation $U$ of $%
\left\{ \xi _{k}\right\} $ into the canonical basis $\left\{ |k\rangle
\right\} \subset \ell ^{2}\left( \mathbf{N}\right) $ is real in the sense $%
\bar{U}:=CUJ=U$, and thus $\tilde{U}:=CU^{\dagger }J=U^{\dagger }$. Hence
amplitude operator $\chi :\mathcal{G}\rightarrow \mathcal{F}\otimes \mathcal{%
H}$ which was defined above by the transposition of $\upsilon U^{\dagger
}=\upsilon \tilde{U}\equiv \tilde{\chi}$, is equivalent to $\tilde{\upsilon}$%
: $\chi =\left( U\otimes I\right) \tilde{\upsilon}$. Thus 
\begin{equation*}
J\mathbb{\chi }^{\dagger }\mathbb{\chi }J=\mathrm{Tr}_{\mathcal{H}}\upsilon
\upsilon ^{\dagger }=\rho ,\quad \mathrm{Tr}_{\mathcal{F}}\chi \chi
^{\dagger }=\mathrm{Tr}_{\mathcal{G}}\upsilon \upsilon ^{\dagger }=\sigma .
\end{equation*}
Moreover, it satisfies the conditions (\ref{1.3}) since $\omega =\upsilon
\upsilon ^{\dagger }\in \mathcal{A}\otimes \mathcal{B}$: 
\begin{equation*}
J\chi ^{\dagger }\left( I\otimes B\right) ^{\dagger }\chi J=\mathrm{Tr}_{%
\mathcal{H}}\left( I\otimes B\right) \omega \in \mathcal{A},\;\mathrm{Tr}_{%
\mathcal{F}}\chi \tilde{A}\chi ^{\dagger }=\mathrm{Tr}_{\mathcal{G}}\left(
A\otimes I\right) \omega \in \mathcal{B}.
\end{equation*}

The uniqueness up to the $U$ follows from the obvious isometricity of the
families 
\begin{equation*}
\left\{ \sum_{k}|k\rangle \eta ^{\dagger }\psi _{k}\left( n\right) :n\in 
\mathbb{N},\eta \in \mathcal{H}\right\} ,\quad \left\{ \sum_{k}\eta
^{\dagger }\psi _{k}\left( n\right) \xi _{k}^{\dagger }:n\in \mathbb{N},\eta
\in \mathcal{H}\right\}
\end{equation*}
of vectors $\left( I\otimes \eta ^{\dagger }\right) \chi |n\rangle $ in $%
\mathcal{F}\subseteq \ell ^{2}\left( \mathbf{N}\right) $ and of $\left(
\langle n|\otimes \eta ^{\dagger }\right) \upsilon $ in $\mathcal{F}%
^{\dagger }$ which follows from 
\begin{equation*}
\mathrm{Tr}_{\mathcal{G}}|n\rangle \langle m|\chi ^{\dagger }\left( I\otimes
\eta \eta ^{\dagger }\right) \chi =\mathrm{Tr}_{\mathcal{F}}\upsilon
^{\dagger }\left( |m\rangle \langle n|\otimes \eta \eta ^{\dagger }\right)
\upsilon .
\end{equation*}
Thus they are unitary equivalent in the minimal space $\mathcal{F}$. So the
entangling operator $\chi $ is defined in the minimal $\mathcal{F}$ up to
unitary equivalence corresponding to the unitary operator $U$ in $\mathcal{F}
$ intertwining the involutions $C$ and $J$.%
\endproof%
%

Note that the entangled state (\ref{1.4}) is written as 
\begin{equation*}
\varpi \left( A\otimes B\right) =\mathrm{Tr}_{\mathcal{H}}B\pi ^{\ast
}\left( A\right) =\mathrm{Tr}_{\mathcal{G}}A\pi \left( B\right) ,
\end{equation*}
where the operator $\pi ^{\ast }\left( A\right) =\mathrm{Tr}_{\mathcal{F}%
}\chi \tilde{A}\chi ^{\dagger }\in \mathcal{B}$, bounded by $\left\|
A\right\| \sigma \in \mathcal{B}_{\ast }$, is in the predual space $\mathcal{%
B}_{\ast }$ for any $A\in \mathcal{L}\left( \mathcal{G}\right) $, and 
\begin{equation}
\pi \left( B\right) =J\chi ^{\dagger }\left( I\otimes B^{\dagger }\right)
\chi J=\widetilde{\chi ^{\dagger }\left( I\otimes B\right) \chi }
\label{1.6}
\end{equation}
is in $\mathcal{A}_{\ast }$ as a trace-class operator in $\mathcal{G}$,
bounded by $\left\| B\right\| \rho \in \mathcal{A}_{\ast }$. The linear map $%
\pi $ is written in the Steinspring form \cite{Sti55} of the normal
completely positive map $B\mapsto \widetilde{\pi \left( B\right) }$, while $%
\pi ^{\ast }:\mathcal{A}\rightarrow \mathcal{B}_{\ast }$ is written in the
Kraus form \cite{Kra71} of the normal CP map $A\mapsto \pi ^{\ast }\left( 
\tilde{A}\right) $ in the canonical orthonormal basis $|k\rangle $ of $%
\mathcal{F}\subseteq \ell ^{2}\left( \mathbb{N}\right) $: 
\begin{equation*}
\pi ^{\ast }\left( A\right) =\sum_{k}\left( \langle k|\otimes I\right) \chi 
\tilde{A}\chi ^{\dagger }\left( |k\rangle \otimes I\right) .
\end{equation*}

A linear map $\pi :\mathcal{B}\rightarrow \mathcal{A}$ is called completely
positive (CP) if the operator matrix 
\begin{equation*}
\pi \left( \mathbf{B}\right) =\pi \left( \left[ B_{ik}\right] \right) =\left[
\pi \left( B_{ik}\right) \right]
\end{equation*}
is positive (in the sense of non-negative definiteness) for every positive
operator-matrix $\mathbf{B}=\left[ B_{ik}\right] $ (which is thus Hermitian, 
$B_{ik}^{\dagger }=B_{ki}$). But the defined in (\ref{1.6}) $\dagger $-map $%
\pi \left( B^{\dagger }\right) =\pi \left( B\right) ^{\dagger }$ is not
necessarily CP but \emph{tilde-completely positive} (TCP) in the sense that
the map $\mathbf{B}\mapsto \pi ^{\symbol{126}}\left( \mathbf{B}\right) $
given by the transposed operator-matrix 
\begin{equation*}
\pi ^{\symbol{126}}\left( \mathbf{B}\right) :=\left[ J\pi \left(
B_{ik}\right) ^{\dagger }J\right] =\left[ J\pi \left( B_{ki}\right) J\right]
\equiv \pi ^{-}\left( \mathbf{B}^{\prime }\right)
\end{equation*}
is positive (in the sense of non-negative definiteness) for every positive
operator-matrix $\left[ B_{ik}\right] =\left[ B_{ki}^{\dagger }\right] $.
Obviously every tilde-positive $\pi :\mathcal{B}\rightarrow \mathcal{A}%
_{\ast }$ as $\pi ^{-}\left( B\right) =J\pi \left( B\right) J$ is positive
for every positive $B$, but it is not necessarily CP even if it is TCP
unless $\mathcal{A}$ (or $\mathcal{B}$) is Abelian. The TCP maps can be
obtained simply by partial tracing 
\begin{equation}
\pi ^{\ast }\left( A\right) =\mathrm{Tr}_{\mathcal{G}}\left( \left( A\otimes
I\right) \omega \right) ,\quad \pi \left( B\right) =\mathrm{Tr}_{\mathcal{H}%
}\left( \left( I\otimes B\right) \omega \right) .  \label{1.8}
\end{equation}
in terms of the compound density operator $\omega =\upsilon \upsilon
^{\dagger }$ for the entangled state 
\begin{equation*}
\varpi \left( A\otimes B\right) =\mathrm{Tr}\left( A\otimes B\right) \omega .
\end{equation*}

\begin{definition}
The normal TCP map $\pi :\mathcal{B}\rightarrow \mathcal{A}_{\ast }$ (and
its dual map $\pi ^{\ast }:\mathcal{A}\rightarrow \mathcal{B}_{\ast }$)
normalized to a probability operator $\rho =\pi \left( I\right) $ as $%
\mathrm{Tr}_{\mathcal{G}}\pi \left( I\right) =1$ (to $\sigma =\pi ^{\ast
}\left( I\right) $ as $\mathrm{Tr}_{\mathcal{H}}\pi ^{\ast }\left( I\right)
=1$) is called coupling of the state $\varsigma $ on $\mathcal{B}$ to $%
\varrho $, (or generalized entanglement of the state $\varrho $ on $\mathcal{%
A}$ to $\varsigma $). The coupling $\pi $ (or entanglement $\pi ^{\ast }$)
is called truly quantum if it is not CP, i.e. if there exists a positive
operator-matrix $\mathbf{B}=\left[ B_{ik}\right] $ with $B_{ik}\in \mathcal{B%
}$ for which $\pi \left( \mathbf{B}\right) =\left[ \pi \left( B_{ik}\right) %
\right] $ is not positive. The copling (entanglement) $\pi =\pi ^{q}=\pi
^{\ast }$ by 
\begin{equation}
\pi ^{q}\left( B\right) =\sigma ^{1/2}\tilde{B}\sigma ^{1/2},\quad B\in 
\mathcal{B}  \label{1.5}
\end{equation}
of the state $\varrho =\varsigma $ on the algebra $\mathcal{A}=\mathcal{B}$
is called standard for the system $\left( \mathcal{B},\varsigma \right) $.
\end{definition}

Note that the standard entanglement is true as soon as the reduced algebra $%
\mathcal{B}_{\sigma }=E_{\sigma }\mathcal{B}E_{\sigma }$ on the support $%
\mathcal{E}_{\sigma }=E_{\sigma }\mathcal{H}$ of the state $\varsigma $ is
not Abelian. (Here $E_{\sigma }$ is the minimal orthoprojector $E\in 
\mathcal{A}$ with $\varsigma \left( E\right) =1$.) In the case of the simple
algebra $\mathcal{B}=\mathcal{L}\left( \mathcal{H}\right) $ it is obvious as 
$\pi ^{q}$ restricted to $\mathcal{B}_{\sigma }$ is the composition of the
nondegenerated multiplication $\mathcal{B}_{\sigma }\ni B\mapsto \sigma
^{1/2}B$ $\sigma ^{1/2}$ (which is CP) and the transposition $\tilde{B}%
=JB^{\dagger }J$ on $\mathcal{B}_{\sigma }$ (which is TCP but not CP if $%
\dim \mathcal{E}_{\sigma }>1$).

The \emph{standard compound state} 
\begin{equation*}
\varpi _{q}\left( A\otimes B\right) =\mathrm{Tr}_{\mathcal{H}}B\sigma ^{1/2}%
\tilde{A}\sigma ^{1/2}=\mathrm{Tr}_{\mathcal{H}}A\sigma ^{1/2}\tilde{B}%
\sigma ^{1/2}
\end{equation*}
on the algebra $\mathcal{B}\otimes \mathcal{B}$ is pure in this case, given
by the amplitude $\upsilon \simeq |\sigma ^{1/2}\rangle \equiv \psi $, where 
$|\sigma ^{1/2}\rangle =\tilde{\chi}$ with $\chi =\sigma ^{1/2}$ and $\tilde{%
\chi}$ defined in (\ref{1.7}) as $\left( \zeta \otimes \eta \right)
^{\dagger }\tilde{\chi}=\eta ^{\dagger }\chi J\zeta $. In particular, any
pure compound state is truly entangled if $\mathrm{rank}\rho =\mathrm{rank}%
\sigma $ is not one because $\pi ^{\ast }\left( A\right) =\chi \tilde{A}\chi
^{\dagger }$ can be decomposed as 
\begin{equation*}
\chi A\chi ^{\dagger }=J\sigma ^{1/2}U^{\dagger }JAJU\sigma ^{1/2}J=\pi
^{q}\left( U^{\dagger }\tilde{A}U\right) ,
\end{equation*}
where $U:\sigma ^{1/2}J\eta \mapsto J\chi ^{\dagger }\eta $ is a unitary
operator from $\mathcal{E}_{\sigma }$ onto the support of $\rho $ in $%
\mathcal{G}$ with nonabelian $\mathcal{B}_{\sigma }=U^{\dagger }\mathcal{A}U=%
\mathcal{L}\left( \mathcal{E}_{\sigma }\right) $.

In the general case of a discretely decomposable $\left( \mathcal{B}%
,\varsigma \right) $ with the density operator $\sigma =\oplus \sigma \left(
i\right) $ having more than one components $\sigma \left( i\right) =\sigma
_{i}p\left( i\right) $ with nonzero probability $p\left( i\right) =\mathrm{Tr%
}\sigma \left( i\right) $ and positive trace one $\sigma _{i}\in \mathcal{T}%
\left( \mathcal{H}_{i}\right) $, the standard compound state is mixed,
described by the decomposable density operator 
\begin{equation}
\omega _{q}=\oplus _{i,j}p\left( j\right) \delta _{j}^{i}|\sigma
_{j}^{1/2}\rangle \langle \sigma _{j}^{1/2}|,\quad A,B\in \mathcal{B}
\label{1.9}
\end{equation}
with zero components $\omega _{q}\left( i,j\right) =\delta _{j}^{i}p\left(
j\right) \omega _{j}$ at $i\neq j$, corresponding to the pure compound
states $\omega _{j}=\psi _{j}\psi _{j}^{\dagger }$. The amplitudes $\upsilon
_{j}\simeq |\sigma _{j}^{1/2}\rangle \equiv \psi _{j}\in \mathcal{H}%
_{j}\otimes \mathcal{H}_{j}$ define the orthogonal decomposition 
\begin{equation*}
\upsilon _{q}=\oplus _{i,j}p\left( j\right) ^{1/2}|\sigma _{j}^{1/2}\rangle
\delta _{j}^{i}\langle i|=\oplus _{ij}\psi \left( i\right) \delta
_{j}^{i}\langle i|
\end{equation*}
of the standard amplitude operator $\upsilon _{q}:\mathcal{F}\rightarrow
\oplus \mathcal{H}_{i}\otimes \mathcal{H}_{j}$ on $\mathcal{F}=\ell
^{2}\left( \mathbb{N}\right) $ with the components 
\begin{equation*}
\upsilon _{q}\left( i,j\right) =\upsilon _{q}\left( j\right) \delta
_{j}^{i},\quad \upsilon _{q}\left( j\right) =\psi \left( j\right) \langle i|,
\end{equation*}
where $\psi \left( j\right) =p\left( j\right) ^{1/2}\psi _{j}$. It
corresponds to the block-diagonal entangling operator $\chi =\left[ \chi
\left( j\right) \delta _{j}^{i}\right] $ with 
\begin{equation*}
\chi \left( j\right) =|j\rangle \otimes \sigma \left( j\right) ^{1/2}=\tilde{%
\upsilon}_{q}\left( j\right) .
\end{equation*}

The so called \emph{separable} compound states, which are given by convex
combinations 
\begin{equation*}
\varpi _{c}\left( A\otimes B\right) =\sum_{n}\varrho _{n}\left( A\right)
\varsigma _{n}\left( B\right) \mu \left( n\right)
\end{equation*}
of the product states $\varrho _{n}\otimes \varsigma _{n}$, are obviously
not true entangled as the corresponding map 
\begin{equation}
\pi ^{c}\left( B\right) =\sum_{n}\varsigma _{n}\left( B\right) \rho _{n}\mu
\left( n\right)  \label{1.10}
\end{equation}
is both CP and TCP$.$

\section{Quantum Entropy via Entanglements}

As we shall prove in this section, the most informative for a quantum system 
$\left( \mathcal{B},\varsigma \right) $ is the standard entanglement $\pi
_{q}=\pi ^{q}=\pi _{q}^{\ast }$ to the probe system $\left( \mathcal{A}%
^{0},\varrho _{0}\right) =\left( \mathcal{B},\varsigma \right) $, described
in (\ref{1.5}).

Let us consider the entangled information and quantum entropies of states by
means of the entangled compound states. To define the quantum information,
we need to apply a quantum version of the relative entropy to compound state
on the algebra $\mathcal{A}\otimes \mathcal{B}$. In classical information
theory the relative entropy is defined as the expectation of the logarithm
of the derivation of the state $\varpi $ with respect to a reference measure 
$\varphi $. The relative entropy measures the information divergence of the
state $\varpi $ with respect to $\varphi $. This information divergence is
equal to the expectation of $\ln \omega \phi ^{-1}=\ln \omega -\ln \phi $ in
the state $\varpi $, where $\omega $ and $\phi $ are the densities of $%
\varpi $ and $\varphi $ with respect to Lebesgue or any other appropriate
measure (with respect to which $\phi $ is invertible, $\phi ^{-1}\phi
=I=\phi \phi ^{-1}$). Such defined, it should be better called the relative
information rather than entropy: indeed many authors reserve the term
relative entropy for the expectation of $-\ln \omega \phi ^{-1}$ such that
it coincides with Boltzmann entropy if $\phi =1$.

In quantum case, however, 
\begin{equation*}
\ln \omega \phi ^{-1}\neq \ln \omega -\ln \phi \neq \ln \phi ^{-1}\omega
\end{equation*}
for noncommuting density operators $\omega $ and $\phi $, thus giving
different possibilities for definition of relative entropy for quantum state 
$\varpi $ with respect to $\varphi $. In \cite{Be_Ohy98, Be_Ohy00} we
investigated the possibility for entangled mutual information based on the
most common definition of quantum relative entropy (information) as the
quantum expectation of the difference $\ln \omega -\ln \phi $, but it was
hard to prove the additivity of the corresponding estimate for quantum
capacity of the nontrivial channels.

Here we take another choice 
\begin{equation}
\mathsf{R}\left( \varpi :\varphi \right) =\mathrm{Tr}\omega \ln \omega
^{1/2}\phi ^{-1}\omega ^{1/2}  \label{4.1}
\end{equation}
suggested in \cite{Be_Sta82, Be_Sta84} for quantum relative entropy of the
state $\varpi $ on an algebra $\mathcal{M}$ with respect to a weight $%
\varphi $, given by a positive invertible operator $\phi \in \mathcal{M}$.
Note that this quantum information divergence is well defined as 
\begin{equation*}
\mathsf{R}\left( \varpi :\varphi \right) =\mathrm{Tr}\phi \eta \left( \phi
^{-1}\omega \right) =\mathrm{Tr}_{\mathcal{F}}\upsilon ^{\dagger }\upsilon
\ln \upsilon ^{\dagger }\phi ^{-1}\upsilon \text{,}
\end{equation*}
where $\eta \left( x\right) =x\ln x$, $\omega =\upsilon \upsilon ^{\dagger }$
and 
\begin{equation*}
\eta \left( \phi ^{-1}\omega \right) :=\phi ^{-1/2}\eta \left( \phi
^{-1/2}\omega \phi ^{-1/2}\right) \phi ^{1/2}
\end{equation*}
as soon as the quantum Radon-Nicodim derivative $\phi ^{-1/2}\omega \phi
^{-1/2}$ of the state $\varpi $ with respect to $\varphi $ is defined as a
positive operator in the Hilbert space. This definition can be extended to
any state $\varpi $ absolutely continuous with respect to $\varphi $ \cite
{Be_Sta86}, i.e. if $\varpi \left( E\right) =0$ for the maximal
null-orthoprojector $E\phi =0$ (otherwise the entropy is infinite by
definition). As proved in \cite{OP93}, it gives a larger relative entropy
than the expectation of $\ln \omega -\ln \phi $, and it has a positive value 
$\mathsf{R}\left( \varpi :\varphi \right) \in \lbrack 0,\infty ]$ if the
states are equally normalized, say (as usually) $\mathrm{Tr}\omega =1=%
\mathrm{Tr}\phi $.

The most important property of the information divergence $\mathsf{R}$ is
its monotonicity property \cite{Uhl77}, i. e. nonincrease of the divergence $%
\mathsf{R}\left( \varpi _{0}:\varphi _{0}\right) $ after the application of
the pre-dual $\mathrm{K}_{\ast }$ of a normal completely positive unital map 
$\mathrm{K}:\mathcal{M}\rightarrow \mathcal{M}^{0}$ to the state $\varpi
_{0} $ and $\varphi _{0}$ on a von Neumann algebra $\mathcal{M}^{0}$: 
\begin{equation}
\varpi =\varpi _{0}\mathrm{K},\varphi =\varphi _{0}\mathrm{K}\Rightarrow 
\mathsf{R}\left( \varpi :\varphi \right) \leq \mathsf{R}\left( \varpi
_{0}:\varphi _{0}\right) .  \label{4.4}
\end{equation}
A quantum statistical morphism $\mathrm{K}$ can only decrease their
information divergence; it can even be made zero by $\varpi _{0}\mathrm{K}%
=\varphi _{0}\mathrm{K}.$

Let $\pi :\mathcal{B}\rightarrow \mathcal{A}_{\ast }$ be an entanglement of
the state $\varrho $ corresponding to the density operator $\rho =\pi \left(
I\right) $. We shall define the $\emph{entangled}$\emph{\ entropy }(or,
better, entangled information) $\mathsf{E}\left( \pi \right) $ as the
relative entropy (information) (\ref{4.1}) of the achieved compound state $%
\varpi $ on $\mathcal{M}=\mathcal{A}\otimes \mathcal{B}$ with respect to the
weight $\varphi =\varrho \otimes \mathrm{Tr}$ corresponding to the density
operator $\phi =\rho \otimes I$: 
\begin{equation}
\mathsf{E}\left( \pi \right) =\mathrm{Tr\,}\omega \ln \omega ^{1/2}\left(
\rho \otimes I\right) ^{-1}\omega ^{1/2}=\mathrm{Tr}_{\mathcal{F}}\upsilon
^{\dagger }\upsilon \ln \upsilon ^{\dagger }\left( \rho \otimes I\right)
^{-1}\upsilon ,  \label{4.2}
\end{equation}
where $\rho ^{-1}$ is quasiinverse to $\rho $ in the case of $\mathrm{rank}%
\rho \neq \mathrm{\dim }\mathcal{G}$.

If $\omega =\psi \psi ^{\dagger }$ is one dimensional orthoprojector
(corresponding to a pure state on the decomposable algebra $\mathcal{A}%
\otimes \mathcal{B}$ with $\upsilon \simeq \psi \in \mathcal{G}\otimes 
\mathcal{H}$), then $\upsilon ^{\dagger }\upsilon =1$, and 
\begin{equation*}
\upsilon ^{\dagger }\left( \rho \otimes I\right) ^{-1}\upsilon =\mathrm{Tr}%
\chi \rho ^{-1}\chi ^{\dagger }=\mathrm{Tr}_{\mathcal{G}}\rho ^{1/2}\rho
^{-1}\rho ^{1/2}=\mathrm{rank}\rho .
\end{equation*}
In this case $\mathsf{E}\left( \pi \right) =\ln \mathrm{rank}\rho \geq 0$,
where $\mathrm{rank}\rho =\mathrm{rank}\sigma $ is the dimensionality of the
range $\mathrm{ran}\chi $ or $\mathrm{ran}\chi ^{\dagger }$. Thus for
quantum entanglements $\pi $ corresponding to pure $\omega $ but mixed $\rho
=\chi ^{\dagger }\chi $ and $\sigma =\chi \chi ^{\dagger }$ (with $\mathrm{%
rank}>1$) the entangled entropy is strictly positive. However it might be
negative as it is in the case of an Abelian $\mathcal{A}$ when 
\begin{equation}
\pi \left( B\right) =\sum |n\rangle \varsigma _{n}\left( B\right) \langle
n|\mu \left( n\right) \equiv \pi ^{d}\left( B\right) .  \label{4.7}
\end{equation}
In this case $\omega =\sum |n\rangle \sigma \left( n\right) \langle n|$, $%
\rho =\sum |n\rangle \mu \left( n\right) \langle n|$, and $\mathsf{E}\left(
\pi \right) =-$\QTR{cal}{\textsf{S}}$\left( \pi \right) $, where 
\begin{equation*}
\mathcal{\mathsf{S}}\left( \pi \right) =-\mathrm{Tr}\omega \ln \frac{\omega 
}{\rho \otimes I}=\sum \mu \left( n\right) \mathcal{\mathsf{S}}\left( \frac{%
\varsigma \left( n\right) }{\mu \left( n\right) }\right) \geq 0
\end{equation*}
is the mean of the conditional von Neumann entropy 
\begin{equation*}
\mathcal{\mathsf{S}}\left( \varsigma _{n}\right) =-\mathrm{Tr}\sigma _{n}\ln
\sigma _{n}
\end{equation*}
on $\mathcal{B}$ corresponding to $\sigma _{n}=\sigma \left( n\right) /\mu
\left( n\right) $. We can call the (mixed) state $\omega $ \emph{essentially
disentangled} (or separable) if $\mathsf{E}\left( \pi \right) \leq 0$; if $%
\omega $ is pure and $\mathsf{E}\left( \pi \right) >0$, it is truly
entangled as the entangling map $\pi $ is obviously not CP in this case.

The total \emph{entangled information} $\mathsf{I}\left( \pi \right) $\ is
defined as the sum of the von Neumann entropy 
\begin{equation}
\mathcal{\mathsf{S}}\left( \varsigma \right) =-\mathrm{Tr}\sigma \ln \sigma
\label{4.5}
\end{equation}
corresponding to the state $\varsigma \left( B\right) =\mathrm{Tr}\pi \left(
B\right) $ and the entangled entropy (information) $\mathsf{E}\left( \pi
\right) $: 
\begin{equation}
\mathsf{I}\left( \pi \right) =\mathrm{Tr\,}\omega \left( \ln \left( \rho
\otimes I\right) ^{-1}\omega -\ln \left( I\otimes \sigma \right) \right) .
\label{4.3}
\end{equation}
Note that $\mathsf{I}\left( \pi \right) \geq 0$, but in general $\mathsf{I}%
\left( \pi \right) \neq \mathsf{I}\left( \pi ^{\ast }\right) $ unless $\rho
\otimes I$ commutes with $\omega $ as in the case of d-entanglement or
Abelian $\mathcal{A}$ when the total information $\mathsf{I}\left( \pi
\right) =\mathsf{I}\left( \pi ^{\ast }\right) $ coincides with the mutual
information $\mathsf{I}_{\mathcal{A};\mathcal{B}}\left( \pi \right) =\mathsf{%
I}_{\mathcal{B};\mathcal{A}}\left( \pi ^{\ast }\right) $ defined in \cite
{Be_Ohy98, Be_Ohy00}. In the classical case when both algebras $\mathcal{A}$
and $\mathcal{B}$ are Abelian, $\mathsf{I}\left( \pi \right) $ coincides
with the Shannon mutual information $\mathsf{I}_{\mathcal{A};\mathcal{B}%
}\left( \pi \right) $.

The following proposition follows from the monotonicity property (\ref{4.4})
of the relative entropy on $\mathcal{M}=\mathcal{A}\otimes \mathcal{B}$ with
respect to the predual $\mathrm{K}_{\ast }\left( \omega _{0}\right) =\varpi
_{0}\left( \mathrm{K}\otimes \mathrm{I}\right) $ to the ampliation $\mathrm{K%
}\otimes \mathrm{I}$ of a normal completely positive unital map $\mathrm{K}:%
\mathcal{A}\rightarrow \mathcal{A}^{0}$.

\begin{proposition}
Let $\pi :\mathcal{B}\rightarrow \mathcal{A}_{\ast }$ be an entanglement of $%
\left( \mathcal{B},\varsigma \right) $ of a state $\varsigma \left( B\right)
=\mathrm{Tr}\pi \left( B\right) $, $B\in \mathcal{B}$ to $\left( \mathcal{A}%
,\varrho \right) $ with the density operator $\rho =\pi \left( I\right) $,
and $\pi _{0}:\mathcal{A}^{0}\rightarrow \mathcal{B}_{\ast }$ be an
entanglement defining $\pi $ by the composition $\pi ^{\ast }=\pi _{0}%
\mathrm{K}$ with a normal completely positive unital map $\mathrm{K}:%
\mathcal{A}\rightarrow \mathcal{A}^{0}$. Then $\mathsf{E}\left( \pi \right)
\leq \mathsf{E}\left( \pi ^{0}\right) $, where $\pi ^{0}=\pi _{0}^{\ast }$,
and thus $\mathsf{I}\left( \pi \right) \leq \mathsf{I}\left( \pi ^{0}\right) 
$. In particular, for any separable $\pi =\pi ^{c}$ where $\pi ^{c}$ is the
convex combination (\ref{1.10}) with $\sum \varsigma _{n}\mu \left( n\right)
=\varsigma $, there exists a not less informative entanglement $\pi ^{d}:%
\mathcal{B}\rightarrow \mathcal{A}^{0}$ with the same $\varsigma \left(
B\right) =\mathrm{Tr}\pi ^{d}\left( B\right) $ and Abelian $\mathcal{A}^{0}$%
, and the standard entanglement $\pi ^{0}=\pi ^{q}$ to $\varrho
_{0}=\varsigma $ with $\mathcal{A}^{0}=\mathcal{B}$ is the maximal one in
the sense that for any entanglement $\pi $ there exists not less informative
q-entanglement with the same $\left( \mathcal{B},\varsigma \right) $.
\end{proposition}

\proof%
%
The first proposition follows from the monotonicity property (\ref{4.7})
applied to the ampliation $\mathrm{K}\left( A\otimes B\right) =\mathrm{K}%
\left( A\right) \otimes B$ of the CP map $\mathrm{K}$ from $\mathcal{A}%
\rightarrow \mathcal{A}^{0}$ to $\mathcal{A}\otimes \mathcal{B}\rightarrow 
\mathcal{A}^{0}\otimes \mathcal{B}$, with the compound state $\mathrm{K}%
_{\ast }\left( \omega _{0}\right) =\varpi _{0}\left( \mathrm{K}\otimes 
\mathrm{I}\right) $ ($\mathrm{I}$ denotes the identity map $\mathcal{B}%
\rightarrow \mathcal{B}$) corresponding to the entanglement $\pi ^{\ast
}=\pi _{0}\mathrm{K}$ and $\mathrm{K}_{\ast }\left( \phi _{0}\right)
=\varrho \otimes \varsigma $ with $\varrho =\varrho _{0}\mathrm{K}$
corresponding to $\varphi _{0}=\varrho _{0}\otimes \varsigma $.

If $\pi $ is separable entanglement (\ref{1.10}), $\pi _{c}=\pi ^{\ast }$
can be decomposed as 
\begin{equation*}
\pi ^{\ast }\left( A\right) =\sum_{n}\tilde{\varrho}_{n}\left( A\right)
\sigma _{n}\mu \left( n\right) =\pi _{0}\left( \mathrm{K}\left( A\right)
\right) .
\end{equation*}
Here $\mathrm{K}\left( A\right) =\sum |n\rangle \varrho _{n}\left( A\right)
\langle n|$ is a normal unital CP map on $\mathcal{A}$ into the Abelian
algebra $\mathcal{A}^{0}$ of diagonal operators on $\mathcal{G}^{0}=\ell
^{2}\left( \mathbb{N}\right) $, and $\pi ^{0}=\pi _{0}^{\ast }=\pi ^{d}$ is
the diagonalizing entanglement (\ref{4.7}).

The inequality (\ref{4.7}) can be also applied to the standard entanglement
corresponding to the compound state (\ref{1.6}) on $\mathcal{B}\otimes 
\mathcal{B}=\oplus _{i,j}\mathcal{B}\left( i\right) \otimes \mathcal{B}%
\left( j\right) $, where $\mathcal{B}\left( i\right) =\mathcal{L}\left( 
\mathcal{H}_{i}\right) $. It is described by the density operator 
\begin{equation}
\omega _{q}=\oplus _{i,j}\mathrm{P}_{\mathcal{B}\otimes \mathcal{B}}\left(
i,j\right) =\oplus _{i}\psi _{i}\psi _{i}^{\dagger }p\left( i\right) \text{,}
\label{4.6}
\end{equation}
where $\mathrm{P}_{\mathcal{B}\otimes \mathcal{B}}\left( i,j\right) =\delta
_{i}^{j}\omega _{i}p\left( i\right) $ is concentrated on the diagonal $%
\oplus _{i}\mathcal{B}\left( i\right) \otimes \mathcal{B}\left( i\right) $
of $\mathcal{B}\otimes \mathcal{B}$. The amplitudes $\psi _{i}\in \mathcal{H}%
_{i}\otimes \mathcal{H}_{i}$ are defined in (\ref{1.6}) as $\psi
_{i}=|\sigma _{i}^{1/2}\rangle $ by the components $\chi _{0}\left( i\right)
=|i\rangle \otimes \sigma \left( i\right) ^{1/2}$ of the standard entangling
operator $\chi _{0}$ on $\mathcal{G}_{0}=\mathcal{H}$ into $\ell ^{2}(%
\mathbf{N})\otimes \mathcal{H}$. Indeed, any entanglement $\pi ^{\ast
}\left( A\right) =\mathrm{Tr}_{\mathcal{F}}\chi A\chi ^{\dagger }$ as a
normal CP map $\mathcal{A}\rightarrow \mathcal{B}$ normalized to the density
operator $\sigma =\mathrm{Tr}_{\mathcal{F}}\chi \chi ^{\dagger }$ can be
represented as the composition $\pi _{0}\mathrm{K}$ of the standard
entanglement $\pi ^{0}=\pi ^{q}$ on $\left( \mathcal{A}^{0},\varrho
_{0}\right) =\left( \mathcal{B},\varsigma \right) $ and a normal unital CP
map $\mathrm{K}:\mathcal{A}\rightarrow \mathcal{B}$. The CP map $\mathrm{K}$
is defined by $\sigma ^{1/2}\mathrm{K}\left( A\right) \sigma ^{1/2}=\pi
^{\ast }\left( \tilde{A}\right) $. It has the form 
\begin{equation*}
\mathrm{K}\left( A\right) =\mathrm{Tr}_{\mathcal{F}_{-}}X^{\dagger }AX,%
\mathrm{\quad }A\in \mathcal{A}
\end{equation*}
where $X$ is an operator $\mathcal{F}_{-}\otimes \mathcal{H}\rightarrow 
\mathcal{G}$, $\mathrm{Tr}_{\mathcal{F}_{-}}X^{\dagger }X=I$ defining the
entangling operator $\chi =\left( I^{-}\otimes \chi _{0}\right) X^{\dagger }$
for $\pi $. Thus the standard entanglement $\pi ^{q}\left( B\right) =\sigma
^{1/2}\tilde{B}\sigma ^{1/2}$ corresponds to the maximal mutual information.%
\endproof%
%

Note that the supremum of the information gain (\ref{4.3}) over all
c-entanglements to the system $\left( \mathcal{B},\varsigma \right) $ is the
von Neumann entropy (\ref{4.5}). It is achieved on any extreme entanglement $%
\pi ^{0}$ with an Abelian $\mathcal{A}^{0}$, given by a decomposition $%
\varsigma =\sum \varsigma _{n}\mu \left( n\right) $ into pure states $%
\varsigma _{n}$. For example by a Schatten decomposition $\sigma
=\sum_{n}|n\rangle \langle n|\nu \left( n\right) $, corresponding to $%
\varsigma _{n}\left( B\right) =\langle n|B|n\rangle $ and $\mu \left(
n\right) =\nu \left( n\right) $. The maximal value $\ln \,\mathrm{rank}%
\mathcal{B}$ of the von Neumann entropy on the algebra $\mathcal{B}$ is
restricted by $\ln \dim \mathcal{B}$ as $\left( \dim \mathcal{B}\right)
^{1/2}\leq \,\mathrm{rank}\mathcal{B}\leq \dim \mathcal{B}$.

\begin{definition}
The maximal total information 
\begin{equation}
\mathsf{H}\left( \varsigma \right) =\sup_{\pi \left( I\right) =\sigma }%
\mathsf{I}\left( \pi \right) =\mathsf{I}\left( \pi _{q}\right) ,  \label{4.8}
\end{equation}
achieved on $\mathcal{A}^{0}=\mathcal{B}$ by the standard q-entanglement $%
\pi ^{q}\left( B\right) =\sigma ^{1/2}\tilde{B}\sigma ^{1/2}$ for a fixed
state $\varsigma \left( B\right) =\mathrm{Tr}_{\mathcal{H}}B\sigma $, is
called q-entropy of the state $\varsigma $. The difference 
\begin{equation*}
\mathsf{H}\left( \pi \right) =\mathsf{H}\left( \varsigma \right) -\mathsf{I}%
\left( \pi \right)
\end{equation*}
is called the q-conditional entropy on $\mathcal{B}$ with respect to $%
\mathcal{A}$.
\end{definition}

Obviously, $\mathsf{H}\left( \pi \right) $ is positive in contrast to $%
\mathsf{S}\left( \varsigma \right) -\mathsf{I}\left( \pi \right) =-\mathsf{E}%
\left( \pi \right) $ which is positive as the averaged conditional entropy $%
\mathsf{S}\left( \pi \right) $ in the case of Abelian $\mathcal{A}$, but
which can achieve also the negative value 
\begin{equation}
\mathcal{\mathsf{S}}\left( \varsigma \right) -\mathsf{H}\left( \varsigma
\right) =-\ln \dim \sigma  \label{4.9}
\end{equation}
the following theorem states.

\begin{theorem}
Let $\mathcal{B}$ be a discrete decomposable algebra on $\mathcal{H}=\oplus
_{i}\mathcal{H}_{i}$, with the state $\varsigma =\oplus \varsigma
_{i}p\left( i\right) $ given by normal states $\varsigma _{i}$ on $\mathcal{L%
}\left( \mathcal{H}_{i}\right) $, and $\mathcal{C}\subseteq \mathcal{B}$ be
its center with the state $p=\varsigma |\mathcal{C}$ induced by the
probability distribution $p\left( i\right) $. Then the q-entropy is given by 
\begin{equation}
\mathsf{H}\left( \varsigma \right) =\mathsf{S}\left( \varsigma \right)
+\sum_{i}p\left( i\right) \ln \mathrm{rank}\sigma _{i}=\mathsf{H}\left(
p\right) +\sum_{i}\mathsf{H}\left( \varsigma _{i}\right) p\left( i\right) ,
\label{4.10}
\end{equation}
where $\mathsf{H}\left( p\right) =-\sum_{i}p\left( i\right) \ln p\left(
i\right) =\mathsf{S}\left( p\right) $, and 
\begin{equation*}
\mathsf{H}\left( \varsigma _{i}\right) =\ln \mathrm{rank}\sigma _{i}-\mathrm{%
Tr}_{\mathcal{H}_{i}}\sigma _{i}\ln \sigma _{i}=\mathsf{S}\left( \varsigma
_{i}\right) +\ln \mathrm{rank}\sigma _{i}.
\end{equation*}
It is positive, $\mathsf{H}\left( \varsigma \right) \in \lbrack 0,\infty ]$,
and if the reduced algebra $\mathcal{B}_{\sigma }=E_{\sigma }\mathcal{B}%
E_{\sigma }$ is finite dimensional, it is bounded, with the maximal value $%
\mathsf{H}\left( \varsigma ^{\circ }\right) =\ln \dim \mathcal{B}_{\sigma }$%
, achieved on the tracial $\sigma ^{\circ }=\left( \dim \mathcal{E}_{\sigma
}\right) ^{-1}E_{\sigma }$, with $\mathcal{E}_{\sigma }=E_{\sigma }\mathcal{H%
}$.
\end{theorem}

\proof%
%
The q-entropy $\mathsf{H}\left( \varsigma \right) $ is the supremum (\ref
{4.8}) of the mutual information (\ref{4.3}) which is achieved on the
standard entanglement, corresponding to the density operator (\ref{4.6}) of
the standard compound state (\ref{1.6}) with $\mathcal{A}=\mathcal{B}$, $%
\rho =\sigma $. Thus $\mathsf{H}\left( \varsigma \right) =\mathsf{E}\left(
\pi ^{q}\right) +$\QTR{cal}{\textsf{S}}$\left( \varsigma \right) $, where 
\begin{equation*}
\mathsf{E}\left( \pi ^{q}\right) =\sum_{i}\psi \left( i\right) ^{\dagger
}\psi \left( i\right) \ln \psi \left( i\right) ^{\dagger }\left( \sigma
\left( i\right) \otimes I\right) ^{-1}\psi \left( i\right) =\sum_{i}p\left(
i\right) \ln \mathrm{rank}\sigma _{i}
\end{equation*}
as $\sigma \left( i\right) =\sigma _{i}p\left( i\right) $, $\upsilon
_{q}\left( i\right) =\psi \left( i\right) \langle i|$, $\psi \left( i\right)
=|\sigma _{i}^{1/2}\rangle p\left( i\right) ^{1/2}$, and 
\begin{eqnarray*}
\psi \left( i\right) ^{\dagger }\left( \sigma \left( i\right) \otimes
I_{i}\right) ^{-1}\psi \left( i\right) &=&\langle \sigma _{i}^{1/2}|\left(
\sigma ^{-1}\otimes I_{i}\right) |\sigma _{i}^{1/2}\rangle \\
&=&\mathrm{Tr}_{\mathcal{H}}\sigma \left( i\right) ^{1/2}\sigma \left(
i\right) ^{-1}\sigma \left( i\right) ^{1/2}=\mathrm{rank}\sigma _{i}.
\end{eqnarray*}
Decomposing the von Neumann entropy as 
\begin{equation*}
\mathsf{S}\left( \varsigma \right) =\sum_{i}\left( \mathsf{S}\left(
\varsigma _{i}\right) -\ln p\left( i\right) \right) p\left( \iota \right) ,
\end{equation*}
we obtain the corresponding decomposition for q-entropy 
\begin{equation*}
\mathsf{H}\left( \varsigma \right) =\sum_{i}\left( \mathsf{H}\left(
\varsigma _{i}\right) -\ln p\left( i\right) \right) p\left( \iota \right) ,
\end{equation*}
where $\mathsf{H}\left( \varsigma _{i}\right) =\ln \mathrm{rank}\sigma _{i}+%
\mathsf{S}\left( \varsigma _{i}\right) $. Due to $0\leq \mathsf{S}\left(
\varsigma _{i}\right) \leq \ln \mathrm{rank}\sigma _{i}$, each $\mathsf{H}%
\left( \varsigma _{i}\right) $ is positive, and it is bounded by $2\ln 
\mathrm{rank}\sigma _{i}=\ln \dim \mathcal{B}_{i}$, where we took into
account that $\mathcal{B}_{i}=E_{\sigma }\mathcal{L}\left( \mathcal{H}%
_{i}\right) E_{\sigma }=\mathcal{L}\left( \mathcal{E}_{i}\right) $ has the
squared dimensionality $\dim \mathcal{E}_{i}=\mathrm{rank}\sigma _{i}$ of $%
\mathcal{E}_{i}=E_{\sigma }\mathcal{H}_{i}$. This gives the dimensional
bound 
\begin{equation*}
\mathsf{H}\left( \varsigma \right) \leq \ln \sum_{i}\left( \dim \mathcal{E}%
_{i}\right) ^{2}=\ln \dim \mathcal{B}_{\sigma }
\end{equation*}
for the q-entropy of a state on the reduced algebra $\mathcal{B}_{\sigma
}=\oplus \mathcal{B}_{i}$. Actually this boundary is achievable, as well as
dimensional capacity 
\begin{eqnarray*}
\mathsf{C}_{q}\left( \mathcal{B}\right) &=&\sup_{p}\sum_{i}p\left( i\right)
\left( 2\sup_{\varsigma _{i}}\mathcal{\mathsf{S}}\left( \varsigma
_{i}\right) -\ln p\left( i\right) \right) \\
&=&-\inf_{p}\sum_{i}p\left( i\right) \left( \ln p\left( i\right) -2\ln 
\mathrm{\dim }\mathcal{H}_{i}\right) =\ln \dim \mathcal{B}.
\end{eqnarray*}
of the algebra $\mathcal{B}$ (in case of finitedimensional $\mathcal{B}$).
Here we used the fact that the supremum of von Neumann entropies 
\begin{equation*}
\mathcal{\mathsf{S}}\left( \varsigma _{i}\right) =-\sum_{i}\mathrm{Tr}_{%
\mathcal{H}_{i}}\sigma _{i}\ln \sigma _{i}
\end{equation*}
for the simple algebras $\mathcal{B}\left( i\right) =\mathcal{L}\left( 
\mathcal{H}_{i}\right) $ with $\dim \mathcal{B}\left( i\right) =\left( \dim 
\mathcal{H}_{i}\right) ^{2}<\infty $ is achieved on the tracial density
operators $\sigma _{i}=\left( \dim \mathcal{H}_{i}\right) ^{-1}I^{i}\equiv
\sigma _{i}^{\circ }$, and the infimum of the relative entropy 
\begin{equation*}
\mathsf{I}\left( p;p^{\circ }\right) =\sum_{i}p\left( i\right) \left( \ln
p\left( i\right) -\ln p^{\circ }\left( i\right) \right) ,
\end{equation*}
where $p^{\circ }\left( i\right) =\dim \mathcal{B}\left( i\right) /\dim 
\mathcal{B}$, is zero, achieved at $p=p^{\circ }$.%
\endproof%
%

\section{Quantum Channel and its Q-Capacity}

Let $\mathcal{H}_{1}$ be a Hilbert space describing a quantum input system
and $\mathcal{H}$ describe its output Hilbert space. A quantum channel is an
affine operation sending each input state defined on $\mathcal{H}_{1}$ to an
output state defined on $\mathcal{H}$ such that the mixtures of states are
preserved. A \emph{deterministic} (noiseless) quantum channel is defined by
a linear isometry $Y\mathrm{:\mathcal{H}}_{1}\rightarrow \mathrm{\mathcal{H}}
$ with $Y^{\dagger }Y=I^{1}$ ($I^{1}$ is the identify operator in $\mathrm{%
\mathcal{H}}_{1}$) such that each input state vector $\eta _{1}\in \mathrm{%
\mathcal{H}}_{1}$, $\left\| \eta _{1}\right\| =1$, is transmitted into an
output state vector $\eta =Y\eta _{1}\in \mathcal{H}$, $\left\| \eta
\right\| =1$. The orthogonal mixtures $\sigma _{1}=\sum_{n}\sigma _{1}\left(
n\right) $ of the pure input states $\sigma _{1}\left( n\right) =\eta
_{1}\left( n\right) \eta _{1}\left( n\right) ^{\dagger }$ are sent into the
orthogonal mixtures $\sigma =\sum_{n}\sigma \left( n\right) $ of the
corresponding pure states $\sigma \left( n\right) =Y\sigma _{1}\left(
n\right) Y^{\dagger }$.

A \emph{noisy quantum channel} sends pure input states $\varsigma _{1}$ on
the algebra $\mathcal{B}^{1}=\mathcal{L}\left( \mathcal{H}_{1}\right) $ into
mixed ones $\varsigma =\Lambda ^{\ast }\left( \varsigma _{1}\right) $ given
by the predual $\Lambda _{\ast }=\Lambda ^{\ast }|\mathcal{B}_{\ast }^{1}$
to a normal completely positive unital map $\Lambda :\mathcal{B}\rightarrow 
\mathcal{B}^{1}$, 
\begin{equation*}
\Lambda \left( B\right) =\mathrm{Tr}_{\mathcal{F}_{1}}Y^{\dagger }BY,\mathrm{%
\ \quad }B\in \mathrm{\mathcal{B}}
\end{equation*}
where $Y$ is a linear operator from $\mathrm{\mathcal{H}}_{1}\otimes 
\mathcal{F}_{+}$ to $\mathrm{\mathcal{H}}$ with $\mathrm{Tr}_{\mathcal{F}%
_{+}}Y^{\dagger }Y=I$, and $\mathcal{F}_{+}$ is a separable Hilbert space of
quantum noise in the channel. Each input mixed state $\varsigma _{1}$ is
transmitted into an output state $\varsigma =\varsigma _{1}\Lambda $ given
by the density operator 
\begin{equation*}
\Lambda ^{\ast }\left( \sigma _{1}\right) =Y\left( \sigma _{1}\otimes
I^{+}\right) Y^{\dagger }\in \mathcal{B}_{\ast }
\end{equation*}
for each density operator $\sigma _{1}\in \mathcal{B}_{\ast }^{1}$, where $%
I^{+}$ is the identity operator in $\mathcal{F}_{+}$.

The input entanglements $\pi ^{1}=\mathcal{A}\rightarrow \mathcal{B}_{\ast
}^{1}$ dual to $\pi _{1}:\mathcal{B}^{1}\rightarrow \mathcal{A}$ will be
denoted as $\pi ^{1}=\kappa =\pi _{1}^{\ast }$. They define the
quantum-quantum correspondences (q-encodings) of probe systems $\left( 
\mathcal{A},\varrho \right) $ with the density operator $\rho =\kappa ^{\ast
}\left( I^{1}\right) $, to the input $\left( \mathcal{B}^{1},\varsigma
_{1}\right) $ of the channel $\Lambda $ with $\sigma _{1}=\kappa \left(
I\right) $. If $\mathrm{K}:\mathcal{A}\rightarrow \mathcal{A}^{0}$ is a
normal completely positive unital map 
\begin{equation*}
\mathrm{K}\left( A\right) =\mathrm{Tr}_{\mathcal{F}_{-}}X^{\dagger }AX,\quad
A\in \mathcal{A},
\end{equation*}
where $X$ is a bounded operator $\mathcal{F}_{-}\otimes \mathcal{G}%
_{0}\rightarrow \mathcal{G}$ with $\mathrm{Tr}_{\mathcal{F}_{-}}X^{\dagger
}X=I^{0}$, the compositions $\kappa =\pi _{0}^{1}\mathrm{K}$, $\pi =\Lambda
^{\ast }\kappa $ are the entanglements of the probe system $\left( \mathcal{A%
},\varrho \right) $ with the channel input $\left( \mathcal{B}^{1},\varsigma
_{1}\right) $ and to the output $\left( \mathcal{B},\varsigma \right) $ via
this channel. The state $\varrho =\varrho _{0}\mathrm{K}$ is given by 
\begin{equation*}
\mathrm{K}^{\ast }\left( \rho _{0}\right) =X\left( I^{-}\otimes \rho
_{0}\right) X^{\dagger }\in \mathcal{A}_{\ast }
\end{equation*}
for each density operator $\rho _{0}\in \mathcal{A}_{\ast }^{0}$, where $%
I^{-}$ is the identity operator in $\mathcal{F}_{-}$. The resulting
entanglement $\pi =\lambda ^{\ast }\mathrm{K}$ defines the compound state $%
\varpi =\varpi _{01}\left( \mathrm{K}\otimes \Lambda \right) $ on $\mathcal{A%
}\otimes \mathcal{B}$ with 
\begin{equation*}
\varpi _{01}\left( A^{0}\otimes B^{1}\right) =\mathrm{Tr\,\,}\tilde{A}%
^{0}\pi _{1}^{0}\left( B^{1}\right) =\mathrm{Tr\,\,}\upsilon _{01}^{\dagger
}\left( A^{0}\otimes B^{1}\right) \upsilon _{01}
\end{equation*}
on $\mathcal{A}^{0}\otimes \mathcal{B}^{1}$. Here $\upsilon _{01}:\mathcal{F}%
_{01}\rightarrow \mathcal{G}_{0}\otimes \mathrm{\mathcal{H}}_{1}$ is the
amplitude operator, uniquely defined by the input compound state $\varpi
_{01}\in \mathcal{A}_{\ast }^{0}\otimes \mathcal{B}_{\ast }^{1}$ up to a
unitary operator $U^{0}$ on $\mathcal{F}_{01}$, and the effect of the input
entanglement $\kappa $ and the output channel $\Lambda $ can be written in
terms of the amplitude operator of the state $\varpi $ as 
\begin{equation*}
\upsilon =\left( X\otimes Y\right) \left( I^{-}\otimes \upsilon _{01}\otimes
I^{+}\right) U
\end{equation*}
up to a unitary operator $U$ in $\mathcal{F}=\mathcal{F}_{-}\otimes \mathcal{%
F}_{01}\otimes \mathcal{F}_{+}$. Thus the density operator $\omega =\upsilon
\upsilon ^{\dagger }$ of the input-output compound state $\varpi $ is given
by $\varpi _{01}\left( \mathrm{K}\otimes \Lambda \right) $ with the density 
\begin{equation}
\left( \mathrm{K}\otimes \Lambda \right) ^{\ast }\left( \omega _{01}\right)
=\left( X\otimes Y\right) \omega _{01}\left( X\otimes Y\right) ^{\dagger },
\label{3.4}
\end{equation}
where $\omega _{01}=\upsilon _{01}\upsilon _{01}^{\dagger }$.

Let $\mathcal{K}_{q}^{1}$ be the set of all normal TCP maps $\kappa :%
\mathcal{A}\rightarrow \mathcal{B}_{\ast }^{1}$ with any probe algebra $%
\mathcal{A}$, normalized as $\mathrm{Tr}\kappa \left( I\right) =1$, and $%
\mathcal{K}_{q}\left( \varsigma _{1}\right) $ be the subset of all $\kappa
\in \mathcal{K}_{q}^{1}$ with $\kappa \left( I\right) =\varsigma _{1}$. Each 
$\kappa \in \mathcal{K}_{q}^{1}\left( \varsigma _{1}\right) $ can be
decomposed as $\kappa _{0}\mathrm{K}$, where $\kappa _{0}^{\ast }=\pi
_{1}^{q}=\kappa _{0}\in \mathcal{K}_{q}\left( \varsigma _{1}\right) $ is the
standard entanglement on $\left( \mathcal{A}^{0},\varrho _{0}\right) =\left( 
\mathcal{B}^{1},\varsigma _{1}\right) $, and $\mathrm{K}$ is a normal unital
CP map $\mathcal{A}\rightarrow \mathcal{B}^{1}$. Further let $\mathcal{K}%
_{c}^{1}$ be the set of all c-entanglements $\kappa $ described by $\kappa
\left( A\right) =\sum_{n}\tilde{\varrho}_{n}\left( A\right) \sigma
_{1}\left( n\right) $, i.e.. $\kappa ^{\ast }=\pi _{1}^{c}$ are convex
combinations (\ref{2.10}) on $\mathcal{B}_{1}$, and $\mathcal{K}_{c}\left(
\varsigma _{1}\right) $ denotes the subset of $\mathcal{K}_{c}$
corresponding to a fixed $\kappa \left( I\right) =\varsigma _{1}$. Each $%
\kappa \in \mathcal{K}_{c}\left( \varsigma _{1}\right) $ can be represented
as $\kappa =\kappa _{0}\mathrm{K}$, where $\kappa _{0}^{\ast }=\pi _{1}^{d}$
is an extreme d-entanglement of an Abelian $\mathcal{A}^{0}$ to $\left( 
\mathcal{B}^{1},\varsigma _{1}\right) $, by a proper choice of the CP map $%
\mathrm{K}:\mathcal{A}\rightarrow \mathcal{B}^{1}$.

Now, let us maximize the entangled mutual entropy for a given quantum
channel $\Lambda $ (and a fixed input state $\varsigma _{1}$) by means of
the above two types of quantum (true) and classical (not true) entanglements 
$\kappa $. The entangled entropy (\ref{4.2}) was defined in the previous
section by the derivative of the probability operator $\omega $ of the
corresponding compound state $\varpi $ on $\mathcal{A}\otimes \mathcal{B}$
with respect to the density operator $\rho \otimes I$ of the product $%
\varphi =\varrho \otimes \mathrm{Tr}_{\mathcal{H}}$. In each case 
\begin{equation*}
\varpi =\varpi _{01}\left( \mathrm{K}\otimes \Lambda \right) ,\quad \varphi
=\varrho _{0}\mathrm{K}\otimes \mathrm{Tr},
\end{equation*}
where $\mathrm{K}$ is a CP map $\mathcal{A}\rightarrow \mathcal{A}^{0}=%
\mathcal{B}^{1}$, $\varpi _{01}$ is one of the corresponding extreme
compound states $\varpi _{q1}$, $\varpi _{d1}$ on $\mathcal{B}^{1}\otimes 
\mathcal{B}^{1}$, and $\varrho _{0}\left( A^{0}\right) =\varpi _{01}\left(
A^{0}\otimes I^{1}\right) $.

\begin{proposition}
The entangled information achieves the following maximal values 
\begin{equation}
\mathsf{E}_{q}\left( \varsigma _{1},\Lambda \right) :=\sup_{\kappa \in 
\mathcal{K}_{q}\left( \varsigma _{1}\right) }\mathsf{E}\left( \kappa ^{\ast
}\Lambda \right) =\mathsf{E}\left( \pi _{1}^{q}\Lambda \right) ,  \label{3.7}
\end{equation}
\begin{equation*}
\mathsf{E}_{c}\left( \varsigma _{1},\Lambda \right) :=\sup_{\kappa \in 
\mathcal{K}^{c}\left( \varsigma _{1}\right) }\mathsf{E}\left( \kappa ^{\ast
}\Lambda \right) =-\mathsf{S}\left( \varsigma _{1},\Lambda \right) .
\end{equation*}
Here $\mathsf{S}\left( \varsigma _{1},\Lambda \right) $ is the minimal von
Neumann mean conditional entropy 
\begin{equation*}
\mathsf{S}\left( \varsigma _{1},\Lambda \right) =\sup_{\pi _{d}^{1}}\mathsf{S%
}\left( \pi _{1}^{d}\Lambda \right) \equiv \mathsf{S}\left( \pi
_{1}^{o}\Lambda \right) ,
\end{equation*}
which is achieved on an extreme (optimal) diagonalizing map $\pi
_{1}^{o}=\pi _{1}^{d}$ with $\mathrm{Tr}\pi _{1}^{o}=\varsigma _{1}$ for all 
$B\in \mathcal{B}^{1}$. The total entangled information achieves
respectively the following maximal values 
\begin{equation*}
\mathsf{I}_{q}\left( \varsigma _{1},\Lambda \right) :=\sup_{\kappa \in 
\mathcal{K}_{q}\left( \varsigma _{1}\right) }\mathsf{I}\left( \kappa ^{\ast
}\Lambda \right) =\mathsf{S}\left( \varsigma _{1}\Lambda \right) +\mathsf{E}%
\left( \pi _{1}^{q}\Lambda \right) ,
\end{equation*}
\begin{equation*}
\mathsf{I}_{c}\left( \varsigma _{1},\Lambda \right) :=\sup_{\kappa \in 
\mathcal{K}^{c}\left( \varsigma _{1}\right) }\mathsf{I}\left( \kappa ^{\ast
}\Lambda \right) =\mathsf{S}\left( \varsigma _{1}\Lambda \right) -\mathsf{S}%
\left( \pi _{1}^{o}\Lambda \right) .
\end{equation*}
They are ordered as 
\begin{equation}
\mathsf{E}_{q}\left( \varsigma _{1},\Lambda \right) \geq \mathsf{E}%
_{c}\left( \varsigma _{1},\Lambda \right) ,\quad \mathsf{I}_{q}\left(
\varsigma _{1},\Lambda \right) \geq \mathsf{I}_{c}\left( \varsigma
_{1},\Lambda \right)  \label{3.9}
\end{equation}
\end{proposition}

\proof%
%
Owing to the monotonicity 
\begin{equation*}
\mathsf{R}\left( \varpi _{01}\left( \mathrm{K}\otimes \Lambda \right)
:\varrho _{0}\mathrm{K}\otimes \mathrm{Tr}\right) \leq \mathsf{R}\left(
\varpi _{01}\left( \mathrm{I}\otimes \Lambda \right) :\varrho _{0}\otimes 
\mathrm{Tr}\right) ,
\end{equation*}
the supremum of over all $\kappa \in \mathcal{K}_{q}\left( \varsigma
_{1}\right) $ is achieved on the standard entanglement $\mathcal{B}%
_{1}\rightarrow \mathcal{A}^{0}$ given by $\kappa ^{\ast }=\pi
_{1}^{q}\equiv \kappa ^{0}$. Due to the same reason the supremum over all
c-entanglements $\kappa \in \mathcal{K}_{c}\left( \varsigma _{1}\right) $
coincides with the supremum over all normal unital maps $\kappa _{0}$ on an
Abelian $\mathcal{A}^{0}$ satisfying the condition $\kappa _{0}\left(
I^{0}\right) =\sigma _{1}$. However the entangled information $\mathsf{E}%
\left( \pi _{1}^{d}\Lambda \right) $ for the not true entanglements $\kappa
_{0}^{\ast }=\pi _{1}^{d}\equiv \kappa ^{0}$ is not positive, coinciding
with the minus the averaged conditional von Neumann entropy $\mathsf{S}%
\left( \pi _{1}^{d}\Lambda \right) $. The minimum of $\mathsf{S}\left( \pi
_{1}^{d}\Lambda \right) $ over all diagonalizing maps $\pi _{1}^{d}$ is
achieved on an optimal pure d-entanglement $\kappa ^{0}=\pi _{1}^{o}$ on $%
\left( \mathcal{B}^{1},\varsigma _{1}\right) $.

The same arguments apply also for the total informations $\mathsf{I}\left(
\pi \right) =\mathsf{E}\left( \pi \right) +\mathsf{S}\left( \varsigma
\right) $, however the suprema $\mathsf{I}_{q}\left( \varsigma ,\Lambda
\right) $ and $\mathsf{I}_{c}\left( \varsigma ,\Lambda \right) $ can be
obtained now straightforward as $\mathsf{S}\left( \varsigma \right) $ does
not depend on the input entanglements with a fixed $\kappa \left( I\right)
=\sigma _{1}$. The inequalities in (\ref{3.9}) simply follow from $\mathcal{K%
}_{q}\left( \varsigma _{1}\right) \supseteq \mathcal{K}_{c}\left( \varsigma
_{1}\right) $%
\endproof%
%

\begin{definition}
The suprema 
\begin{equation*}
\mathsf{C}_{q}\left( \Lambda \right) :=\sup_{\kappa \in \mathcal{K}_{q}^{1}}%
\mathsf{I}\left( \kappa ^{\ast }\Lambda \right) =\sup_{\varsigma _{1}}%
\mathsf{I}_{q}\left( \varsigma _{1},\Lambda \right) ,\;
\end{equation*}
\begin{equation}
\mathsf{C}_{c}\left( \Lambda \right) :=\sup_{\kappa \in \mathcal{K}_{c}^{1}}%
\mathsf{I}\left( \kappa ^{\ast }\Lambda \right) =\sup_{\varsigma _{1}}%
\mathsf{I}_{c}\left( \varsigma _{1},\Lambda \right) ,\;  \label{3.10}
\end{equation}
are called the q- and c-capacities respectively for the quantum channel
defined by a normal unital CP map $\Lambda :\mathcal{B}\rightarrow \mathcal{B%
}^{1}$.
\end{definition}

Obviously the capacities (\ref{3.10}) satisfy the inequalities 
\begin{equation*}
\mathsf{C}_{c}\left( \Lambda \right) \leq \mathsf{C}_{q}\left( \Lambda
\right) .
\end{equation*}

\begin{theorem}
Let $\Lambda \left( B\right) =Y^{\dagger }BY$ be a unital CP map $\mathcal{B}%
\rightarrow \mathcal{B}^{1}$ describing a quantum deterministic (noiseless)
channel. Then 
\begin{equation*}
\mathsf{I}_{c}\left( \varsigma _{1},\Lambda \right) =\mathcal{\mathsf{S}}%
\left( \varsigma _{1}\right) ,\quad \mathsf{I}_{q}\left( \varsigma
_{1},\Lambda \right) =\mathsf{H}\left( \varsigma _{1}\right) ,
\end{equation*}
and thus in this case 
\begin{equation*}
\mathsf{C}_{c}\left( \Lambda \right) =\ln \mathrm{rank}\mathcal{B}^{1},\quad 
\mathsf{C}_{q}\left( \Lambda \right) =\ln \dim \mathcal{B}^{1}
\end{equation*}
\end{theorem}

\proof%
%
It was proved in the previous section for the case of the identity channel $%
\Lambda =\mathrm{I}$, and thus it is also valid for any isomorphism $\Lambda 
$ described by a unitary operator $Y$. In the case of non-unitary $Y$ we can
use the identity 
\begin{equation*}
\mathrm{Tr\,\,}Y\left( \sigma _{1}\otimes I^{+}\right) Y^{\dagger }\ln
Y\left( \sigma _{1}\otimes I^{+}\right) Y^{\dagger }=\mathrm{Tr\,\,}S\left(
\sigma _{1}\otimes I^{+}\right) \ln S\left( \sigma _{1}\otimes I^{+}\right) ,
\end{equation*}
where $S=Y^{\dagger }Y$. Due to this \QTR{cal}{\textsf{S}}$\left( \varsigma
_{1}\Lambda \right) =-\mathrm{Tr\,\,}S\left( \sigma _{1}\otimes I^{+}\right)
\ln S\left( \sigma _{1}\otimes I^{+}\right) $. However in the case of the
noiseless channel $I^{+}=1$, $S=I$, and thus 
\begin{equation*}
\mathcal{\mathsf{S}}\left( \varsigma \right) =\mathcal{\mathsf{S}}\left(
\varsigma _{1}\Lambda \right) =\mathcal{\mathsf{S}}\left( \varsigma
_{1}\right) .
\end{equation*}
Moreover, as $\upsilon =\left( X\otimes Y\right) \left( I^{-}\otimes
\upsilon _{01}\right) $, $\upsilon ^{\dagger }\upsilon =\left( I^{-}\otimes
\upsilon _{01}\right) ^{\dagger }\left( R\otimes I\right) \left(
I^{-}\otimes \upsilon _{01}\right) =\upsilon _{1}^{\dagger }\upsilon _{1}$,
where $R=X^{\dagger }X$, $\upsilon _{1}=\left( X\otimes I\right) \left(
I^{-}\otimes \upsilon _{01}\right) $, and 
\begin{equation*}
\upsilon ^{\dagger }\left( \rho \otimes I\right) ^{-1}\upsilon =\left(
I^{-}\otimes \upsilon _{01}\right) ^{\dagger }\left( X^{\dagger }\rho
^{-1}X\otimes I\right) \left( I^{-}\otimes \upsilon _{01}\right) =\upsilon
_{1}^{\dagger }\left( \rho \otimes I\right) ^{-1}\upsilon _{1},
\end{equation*}
where $\rho =X\left( I^{-}\otimes \rho _{0}\right) X^{\dagger }$. Hence 
\begin{equation*}
\mathsf{E}\left( \pi _{1}\Lambda \right) =\mathrm{Tr}_{\mathcal{F}}\upsilon
_{1}^{\dagger }\upsilon _{1}\ln \upsilon _{1}^{\dagger }\left( \rho \otimes
I\right) ^{-1}\upsilon _{1}=\mathsf{E}\left( \pi _{1}\right) ,
\end{equation*}
where $\pi _{1}=\kappa ^{\ast }$. Thus $\mathsf{C}_{q}\left( \Lambda \right)
=\sup_{\varsigma _{1}}\mathsf{H}\left( \varsigma _{1}\right) $, $\mathsf{C}%
_{c}\left( \Lambda \right) =\sup_{\varsigma _{1}}\mathsf{S}\left( \varsigma
_{1}\right) $ due to 
\begin{equation*}
\sup_{\kappa \in \mathcal{K}_{q}\left( \varsigma _{1}\right) }\mathsf{E}%
\left( \kappa ^{\ast }\Lambda \right) =\mathsf{H}\left( \varsigma
_{1}\right) ,\sup_{\kappa \in \mathcal{K}_{c}\left( \varsigma _{1}\right) }%
\mathsf{E}\left( \kappa ^{\ast }\Lambda \right) =\mathsf{S}\left( \varsigma
_{1}\right) .
\end{equation*}
Therefore $\mathsf{C}_{q}\left( \Lambda \right) =\ln \dim \mathcal{B}^{1}=%
\mathsf{C}_{q}\left( \mathcal{B}^{1}\right) $, $\mathsf{C}_{c}\left( \Lambda
\right) =\ln \mathrm{rank}\mathcal{B}^{1}=\mathsf{C}_{c}\left( \mathcal{B}%
^{1}\right) $.%
\endproof%
%

In order to consider block entanglements let us introduce the product
systems $\left( \mathcal{B}^{\otimes n},\varsigma ^{\otimes n}\right) $ on
the tensor product $\mathcal{H}^{\otimes n}=\otimes _{l=1}^{n}\mathcal{H}%
_{l} $ of identical spaces $\mathcal{H}_{l}=\mathcal{H}$, and the product
channels $\Lambda _{\ast }^{\otimes n}:\mathcal{B}_{\ast }^{n}\mapsto 
\mathcal{B}_{\ast }^{\otimes n}$, the preduals of the normal unital CP
product maps 
\begin{equation*}
\Lambda ^{\otimes n}:\mathcal{B}^{\otimes n}\rightarrow \mathcal{B}%
^{n}=\otimes _{l=1}^{n}\mathcal{B}_{l}^{1},\quad \Lambda ^{\otimes n}\left(
B^{\otimes n}\right) =\Lambda \left( B\right) ^{\otimes n},
\end{equation*}
where $\mathcal{B}_{l}=\mathcal{B}^{1}$ for all $l$. Obviously 
\begin{eqnarray*}
\mathsf{S}\left( \varsigma ^{\otimes n}\right) &=&n\mathsf{S}\left(
\varsigma \right) ,\;\mathsf{C}_{c}\left( \mathcal{B}^{\otimes n}\right) =n%
\mathsf{C}_{c}\left( \mathcal{B}\right) , \\
\mathsf{H}\left( \varsigma ^{\otimes n}\right) &=&n\mathsf{H}\left(
\varsigma \right) ,\;\mathsf{C}_{q}\left( \mathcal{B}^{\otimes n}\right) =n%
\mathsf{C}_{q}\left( \mathcal{B}\right) .
\end{eqnarray*}
However it not obvious that this additivity should take place for 
\begin{eqnarray*}
\;\mathsf{I}_{c}^{n}\left( \varsigma _{1},\Lambda \right) &=&\mathsf{I}%
_{c}\left( \varsigma _{1}^{\otimes n},\Lambda ^{\otimes n}\right) ,\;\mathsf{%
C}_{c}^{n}\left( \Lambda \right) =\mathsf{C}_{c}\left( \Lambda ^{\otimes
n}\right) , \\
\mathsf{I}_{q}^{n}\left( \varsigma _{1},\Lambda \right) &=&\mathsf{I}%
_{q}\left( \varsigma _{1}^{\otimes n},\Lambda ^{\otimes n}\right) ,\;\mathsf{%
C}_{q}^{n}\left( \Lambda \right) =\mathsf{C}_{q}\left( \Lambda ^{\otimes
n}\right) .
\end{eqnarray*}
the suprema of the mutual information $\mathsf{I}\left( \kappa ^{\ast
}\Lambda ^{\otimes n}\right) $ over the set $\mathcal{K}_{q}\left( \varsigma
_{1}^{\otimes n}\right) $ of all input entanglements $\kappa :\mathcal{A}%
\rightarrow \mathcal{B}_{\ast }^{n}$ with any probe algebra $\mathcal{A}$,
normalized as $\kappa \left( I\right) =\sigma _{1}^{\otimes n}$, and with
any such normalization respectively, $\kappa \in \mathcal{K}_{q}^{n}$. It is
easily seen, by applying the monotonicity property (\ref{4.4}) with respect
to the normal unital CP map $\mathrm{K}:\mathcal{A}\rightarrow \mathcal{A}%
^{0}=\mathcal{B}^{n}$ in the decomposition $\kappa =\pi _{q}^{n}\mathrm{K}$,
where 
\begin{equation*}
\pi _{q}^{n}\left( A^{0}\right) =\left( \sigma _{1}^{\otimes n}\right) ^{1/2}%
\tilde{A}^{0}\left( \sigma _{1}^{\otimes n}\right) ^{1/2},\quad A^{0}\in 
\mathcal{B}^{n},
\end{equation*}
that the quantities $\mathsf{I}_{q}^{n}\left( \varsigma _{1},\Lambda \right) 
$ are additive: 
\begin{equation*}
\mathsf{I}_{q}^{n}\left( \varsigma _{1},\Lambda \right) =n\mathsf{I}%
_{q}\left( \varsigma _{1},\Lambda \right) ,\quad \mathsf{C}_{q}^{n}\left(
\Lambda \right) =n\mathsf{C}_{q}\left( \Lambda \right) \text{.}
\end{equation*}
Note that if the supremum $\sup_{\varsigma _{n}}\mathsf{I}_{q}\left(
\varsigma _{n},\Lambda ^{\otimes n}\right) $ is taken over all states $%
\varsigma _{n}\in \mathcal{B}_{\ast }^{n}$ but not just over $\varsigma
_{1}^{\otimes n}$, it might be possible to achieve more than $n\mathsf{C}%
_{q}\left( \Lambda \right) $. However for the classical entanglements this
additivity cannot be proved even in the case of $\varsigma _{n}=\varsigma
_{1}^{\otimes n}$: 
\begin{equation*}
\mathsf{I}_{c}^{n}\left( \varsigma _{1},\Lambda \right) \geq n\mathsf{I}%
_{c}\left( \varsigma _{1},\Lambda \right) ,\quad \mathsf{C}_{c}^{n}\left(
\Lambda \right) \geq n\mathsf{C}_{c}\left( \Lambda \right) .
\end{equation*}
The superadditivity implies that the quantities $\mathsf{i}_{c}^{n}=\mathsf{I%
}_{c}^{n}/n,\mathsf{c}_{c}^{n}=\mathsf{C}_{c}^{n}/n$ have the limits 
\begin{equation*}
\mathsf{i}_{c}\left( \varsigma _{1},\Lambda \right) =\lim_{n\rightarrow
\infty }\frac{1}{n}\mathsf{I}_{c}^{n}\left( \left( \varsigma _{1},\Lambda
\right) \right) \geq \mathsf{I}_{c}\left( \varsigma _{1},\Lambda \right)
,\quad \mathsf{c}_{c}\left( \Lambda \right) =\lim_{n\rightarrow \infty }%
\frac{1}{n}\mathsf{C}_{c}^{n}\left( \Lambda \right) \geq \mathsf{C}%
_{c}\left( \Lambda \right)
\end{equation*}
which are usually taken as the bounds of classical information and capacity 
\cite{Hol73, StVa78} for a quantum channel $\Lambda $. As it has been
recently proved in \cite{HJS96} under a certain regularity condition, the
upper bound $\mathsf{C}\left( \Lambda \right) $ is indeed asymptotically
achievable by long block classical-quantum encodings. Note that the
c-quantities $\mathsf{i}_{c}^{n}$, $\mathsf{c}_{c}^{n}$ are not easy to
evaluate for each $n$, but they all are bounded by the corresponding
q-quantities $\mathsf{i}_{q}=\frac{1}{n}\mathsf{I}_{q}^{n}=\mathsf{I}_{q}$
and $\mathsf{c}_{q}=\lim \frac{1}{n}\mathsf{C}_{q}^{n}$: 
\begin{equation*}
\mathsf{i}_{c}\left( \varsigma _{1},\Lambda \right) \leq \mathsf{i}%
_{q}\left( \varsigma _{1},\Lambda \right) ,\frak{\quad }\mathsf{c}_{c}\left(
\Lambda \right) \leq \mathsf{c}_{q}\left( \Lambda \right) .
\end{equation*}

In order to measure the ''real'' entangled information of quantum channels
''without the classical part'', another quantity, the ''coherent
information'' was introduced in \cite{Sch93}. It is defined in our notations
as 
\begin{equation*}
\mathsf{I}_{s}\left( \varsigma _{1},\Lambda \right) =\mathsf{I}_{q}\left(
\varsigma _{1},\Lambda \right) -\mathcal{\mathsf{S}}\left( \varsigma \right)
,\quad \mathsf{C}_{s}^{1}\left( \Lambda \right) =\sup_{\varsigma _{1}}%
\mathsf{I}_{s}\left( \varsigma _{1},\Lambda \right)
\end{equation*}
Obviously $\mathsf{I}_{s}\left( \varsigma _{1},\Lambda \right) \leq \mathsf{E%
}\left( \varsigma _{1},\Lambda \right) $, and $\mathsf{C}_{s}^{1}\left(
\Lambda \right) \leq \mathsf{C}_{e}^{1}\left( \Lambda \right)
=\sup_{\varsigma _{1}}\mathsf{E}\left( \varsigma _{1},\Lambda \right) $. The
supremum $\mathsf{C}_{s}^{n}\left( \Lambda \right) =\mathsf{C}_{s}\left(
\Lambda ^{\otimes n}\right) $ of $\mathsf{I}_{s}\left( \varsigma
_{1}^{n},\Lambda ^{\otimes n}\right) $ over all states $\varsigma
_{1}^{n}\in \otimes _{1}^{n}\mathcal{B}_{\ast }^{1}$ in general is not
additive but superadditive, and the coherent capacity is defined as the
limit 
\begin{equation*}
\mathsf{c}_{s}\left( \Lambda \right) =\lim_{n\rightarrow \infty }\frac{1}{n}%
\mathsf{C}_{s}^{n}\left( \Lambda \right) \leq \lim_{n\rightarrow \infty }%
\frac{1}{n}\mathsf{C}_{e}^{n}\left( \Lambda \right) =\mathsf{c}_{e}\left(
\Lambda \right) .
\end{equation*}
Obviously this capacity has the bounds $\mathsf{c}_{s}\left( \Lambda \right)
\leq \mathsf{c}_{q}\left( \Lambda \right) $ as $\mathsf{c}_{e}\left( \Lambda
\right) \leq \mathsf{c}_{q}\left( \Lambda \right) $ due to $\mathsf{E}%
_{s}\left( \varsigma _{1}^{n},\Lambda ^{\otimes n}\right) \leq \mathsf{I}%
_{q}\left( \varsigma _{1}^{n},\Lambda ^{\otimes n}\right) \geq 0$ for each
input state $\varsigma _{1}$.

Thus the entangled information $\mathsf{I}_{q}\left( \varsigma _{1},\Lambda
\right) $ for a single channel corresponding to the standard entanglement $%
\kappa $ gives upper bound of all other informations and is good analog of
the corresponding classical quantities.


\begin{thebibliography}{99}
\bibitem{Be00d}  Belavkin, V.P., Open Sys. and Information Dyn. \textbf{7},
101--129, 2000.

\bibitem{Ben93}  Bennett, C.H. and G. Brassard, C. Cr\'{e}peau, R. Jozsa, A.
Peres, W.K. Wootters, Phys. Rev. Lett., \textbf{70}, pp.1895-1899, 1993.

\bibitem{Sch93}  Schumacher, B., Phys. Rev. A, \textbf{51}, pp.2614-2628,
1993; Phys. Rev. A, \textbf{51}, pp.2738-2747, 1993, Phys. Rev. A, \textbf{54%
}, p.2614, 1996.

\bibitem{JoSch94}  Jozsa, R. and B. Schumacher, J. Mod. Opt., \textbf{41},
pp.2343-2350, 1994.

\bibitem{Be_Ohy98}  Belavkin, V.P. and M. Ohya, Los Alamos Archive
Quant--Ph/9812082, 1-16, 1998.

\bibitem{Be_Ohy00}  Belavkin, V.P. and M. Ohya, Los Alamos Archive
Quant--Ph/0004069, 1-20, 2000.

\bibitem{Lin73}  Lindblad, G., ``Entropy, Information and Quantum
Measurement''. Comm. in Math. Phys. \textbf{33}, p305--322 (1973).

\bibitem{Ara76}  Araki, H., ``Relative Entropy of states of von Neumann
Algebras'', Publications RIMS, Kyoto University, \textbf{11}, pp. 809--833,
(1976).

\bibitem{Ume}  Umegaki, H., Kodai Math. Sem. Rep., \textbf{14}, pp59-85,
1962.

\bibitem{Be_Sta82}  Belavkin, V.P. and P. Staszewski, Annals de l'institut
Henri Poincare: Phys. Theor. \textbf{37}, 51--57, 1982.

\bibitem{Be__Sta84}  Belavkin, V.P. and P. Staszewski, Rep. on Math. Phys. 
\textbf{20}, 373--384, 1984.

\bibitem{Be_Sta86}  Belavkin, V.P. and P. Staszewski, Rep. on Math. Phys. 
\textbf{24}, 49--55, 1986.

\bibitem{OP93}  Ohya, M. and D.Petz, ``Quantum Entropy and Its Use''\textbf{,%
} Springer, 1993.

\bibitem{Sti55}  Stinespring, W. F., Proc. Amer. Math. Soc. \textbf{6}, p.
211 (1955).

\bibitem{Kra71}  Kraus, K., Ann. Phys. \textbf{64}, p. 311 (1971).

\bibitem{Uhl77}  Uhlmann, A., Commun. Math. Phys., \textbf{54}, pp.21-32,
1977.

\bibitem{Hol73}  Holevo, A.S, Probl. Peredachi Inform., \textbf{9}, no. 3,
pp3-11, 1973.

\bibitem{StVa78}  Stratonovich, R.S. and A.G. Vancian, Probl. Control
Inform. Theory, \textbf{7}, no. 3, pp.161-174, 1978.

\bibitem{HJS96}  Hausladen P., R. Jozsa, B. Schumacher, M. Westmoreland, W.
Wootters, Phys. Rev. A, \textbf{54}, no. 3, pp. 1869-1876, 1996.
\end{thebibliography}
\end{document}